\documentclass[twocolumn,amsmath,amssymb,pra,superscriptaddress]{revtex4}
\usepackage{graphicx}
\usepackage{amsmath,amsfonts,amssymb,amsthm}
\usepackage{fancyhdr}
\usepackage{mathrsfs}
\usepackage{epstopdf}
\usepackage{setspace}
\usepackage{bm}
\usepackage[colorlinks]{hyperref}
\usepackage{xcolor}
\usepackage{dcolumn}

\begin{document}

\title{Resonance inversion in a superconducting cavity coupled to artificial atoms and a microwave background}

\pacs{}

\author{Juha Lepp\"akangas}
\affiliation{Physikalisches Institut, Karlsruhe Institute of Technology,  76131 Karlsruhe, Germany}
\affiliation{HQS Quantum Simulations GmbH, 76131 Karlsruhe, Germany}

\author{Jan David Brehm}
\affiliation{Physikalisches Institut, Karlsruhe Institute of Technology,  76131 Karlsruhe, Germany}

\author{Ping Yang}
\affiliation{Physikalisches Institut, Karlsruhe Institute of Technology,  76131 Karlsruhe, Germany}

\author{Lingzhen Guo}
\affiliation{Max Planck Institute for the Science of Light,  91058 Erlangen, Germany}

\author{Michael Marthaler}
\affiliation{HQS Quantum Simulations GmbH, 76131 Karlsruhe, Germany}
\affiliation{Institut f\"ur Theorie der Kondensierten Materie, Karlsruhe Institute of Technology, 76131 Karlsruhe, Germany}
\affiliation{ Theoretische Physik, Universit\"at des Saarlandes, 66123 Saarbr\"ucken, Germany}

\author{Alexey V. Ustinov}
\affiliation{Physikalisches Institut, Karlsruhe Institute of Technology,  76131 Karlsruhe, Germany}
\affiliation{Russian Quantum Center, National University of Science and Technology MISIS, 119049 Moscow, Russia}

\author{Martin Weides}
\affiliation{Physikalisches Institut, Karlsruhe Institute of Technology,  76131 Karlsruhe, Germany}
\affiliation{School of Engineering, University of Glasgow, Glasgow G12 8QQ, UK}

\begin{abstract}
We demonstrate how heating of an environment can invert the line shape of a driven cavity.
We consider a superconducting coplanar cavity coupled to multiple artificial atoms.
The measured cavity transmission is characterized by Fano-type resonances with a shape that is continuously tunable by
bias current through nearby (magnetic flux) control lines.
In particular, the same dispersive shift of the microwave cavity can be observed as a peak or a dip.
We find that this Fano-peak inversion is possible due to a tunable interference between a microwave transmission through a background, with reactive and dissipative properties, and through the cavity, affected by bias-current induced heating.
The background transmission occurs due to crosstalk between the control and transmission lines.
We show how such background can be accounted for by Jaynes-Cummings type models via modified boundary conditions between the cavity and transmission lines.
We find generally that whereas resonance positions determine system energy levels, resonance shapes give information on system  fluctuations and dissipation.
\end{abstract}

\maketitle

\section{Introduction}

A Fano resonance~\cite{Fano1961} is a fundamental effect of wave propagation.
It appears in a wide range of physical systems, including light propagation in photonic devices~\cite{FanoPhotonics2017},
light interaction with nano- and microstructures~\cite{Miroshnichenko2005,Xu2006,Kroner2008,Lukyanchuk2010, Chen2014},
charge transport in nanoelectronics~\cite{FanoNanoStructures2010}, and inelastic scattering of elementary particles~\cite{Fano1961}.
The Fano resonance emerges due to an interference effect between two parallel paths connecting input and output scattering states:
transmission through a continuous-mode or wide background state and transmission through a discrete or narrow energy-level.
The resulting resonance can be both admitting or reflecting (anti-resonance) depending on the details of the system.
As many other interference effects, it has numerous practical applications in metrology and optical engineering~\cite{FanoNanoStructures2010,FanoPhotonics2017}.

In  microwave devices, wanted or unwanted Fano-type resonances can easily emerge, for example, from capacitive or inductive background coupling between different ports of a resonant circuit~\cite{Lv2016}.
In the following, we call this type of coupling a microwave background.
In state-of-the-art superconducting quantum-information devices~\cite{Blais2004,Devoret2013,Gu2017,Krinner2018} a large number of qubits with their control lines are integrated into a small-sized chip,
with possible further size-optimization to reduce decoherence mechanisms, such as non-equilibrium quasiparticle tunneling~\cite{Riwar2016} or field focusing~\cite{Bothner2017}.
A quantum-state measurement in such circuits is commonly based on microwave transmission through readout resonators~\cite{Wallraff2005}.
When increasing circuit complexity and scaling up qubit numbers, it is difficult to avoid multiple interference paths
for microwave signals propagating through the circuit.
It is then also of high interest to understand in detail how Fano resonances can appear in such circuits
and how to account for them in most commonly used theoretical models.

In this article, we investigate microwave transmission across a superconducting coplanar resonator coupled to multiple artificial atoms.
The measured transmission is characterized by Fano-type resonances
with a shape that is continuously tunable by current bias through nearby magnetic-flux control lines.
In particular, we observe that the very same dispersive shift of the cavity can be seen as either a peak or a dip,
depending on the current bias. The experiment is also characterized by large off-resonance transmission.

We investigate the observed effects further by establishing a theoretical model for cavity transmission in the presence of a microwave background.
The background transmission accounts for a crosstalk between the input and output transmission lines through control lines bypassing the coplanar microwave cavity.
We find that the well-known Jaynes- and Tavis-Cummings models~\cite{WallsMilburn,TavisCummings1968,Wallraff2008,Fink2009} of the cavity-atom interaction are valid also for the considered system.
The background transmission can be accounted for by modifying boundary conditions between the cavity and transmission-line microwave fields.

Using the established model, 
we find that dissipation and incoherent transitions can strongly affect the cavity line shape. 
A transformation from a peak to a dip, which we call here a resonance inversion,
is possible through a changing interference between a background microwave transmission, with reactive and dissipative properties,
and a cavity transmission, with dissipative or incoherent dynamics.
A dip appears when the cavity transmission is comparable with the background transmission and dissipation.

Based on the established model, we explain the observed experimental features as an interplay between bias-induced heating and background transmission: The applied DC bias currents cause local heating of the bias leads, which in turn induce incoherent energy-level transitions in the cavity-atom system.
Increasing rates of such transitions, by increasing the bias currents, reduces the cavity transmission and smoothly changes the spectroscopic response of the device from a peak to a dip, the latter appearing when the cavity transmission becomes comparable with the background transmission and dissipation. This allows for the full tunability of the Fano resonance shape.
We also demonstrate how the local temperature of the system as well as average qubit $T_1$ and $T_2$ times
can be estimated from the line-shape of the resonator.

The article is organized as follows. In Sec.~\ref{sec:System}, we introduce our model
of microwave transmission across a coplanar resonator coupled to artificial atoms. We show how microwave propagation beside the cavity can be included by additional terms in the cavity-line boundary conditions. 
In Sec.~\ref{sec:FanoResonance}, we study theoretically the effect of dissipation and decoherence on the form of Fano resonances.
In particular, we show how the line shape of a microwave resonator can change from a peak to a dip
in various situations.
In Sec.~\ref{sec:Experiment}, we present our experimental results and
estimate the local temperature of the cavity-qubit system under different biasing conditions
from the line shape of the resonator.
We also discuss how to estimate average qubit $T_1$ and $T_2$ times from the resonator line shape.
Conclusions and discussion are given in Sec.~\ref{sec:Conclusions}.


\section{System and model}\label{sec:System}
\begin{figure}
\includegraphics[width=\columnwidth]{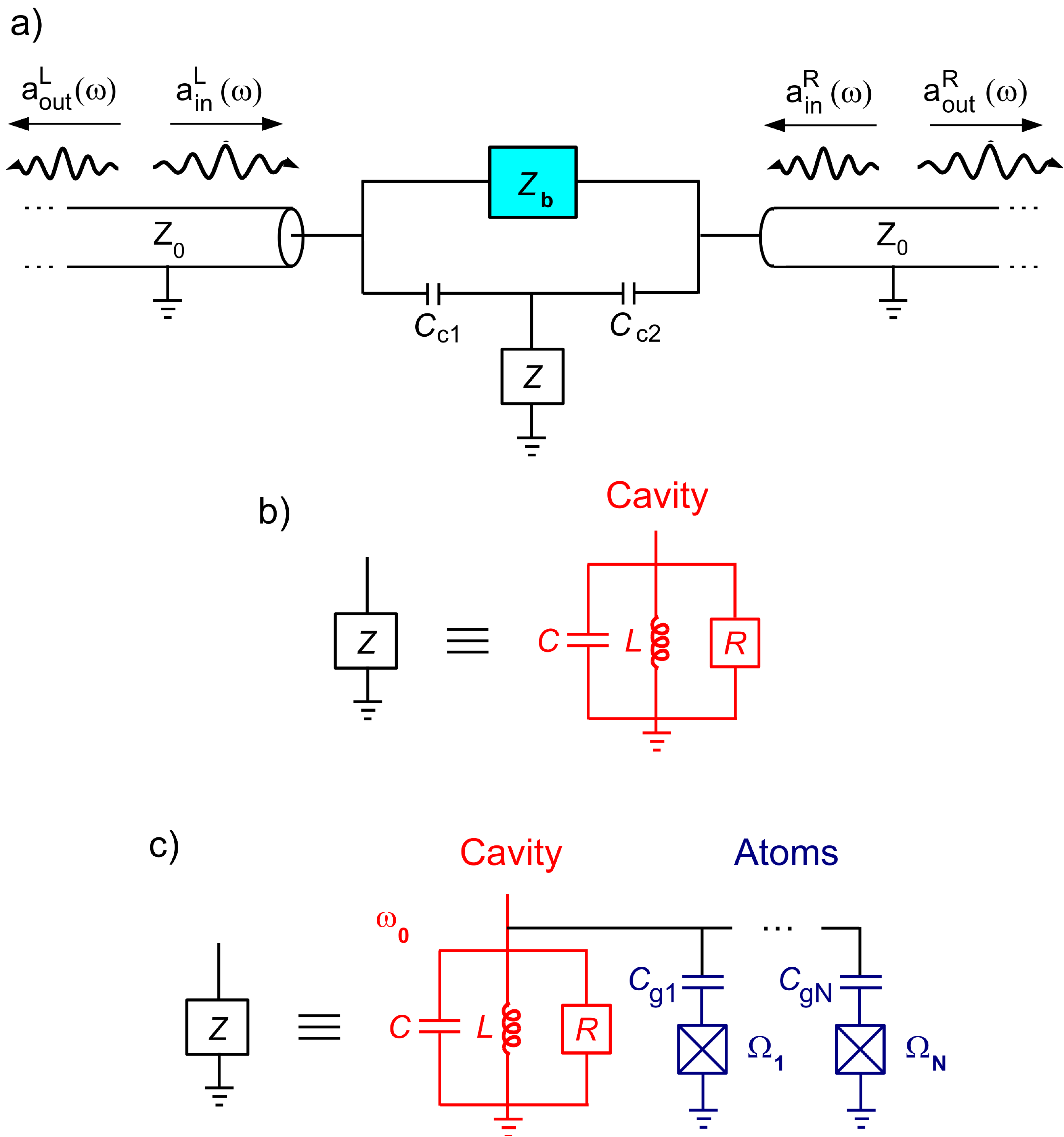}
\caption{
(a) Incoming and outgoing microwave fields propagate in two semi-infinite transmission lines with characteristic impedance $Z_0$. They are connected
via capacitors $C_{\rm c}$ to a microwave resonator modeled by an element $Z$. The transmission lines are also connected directly to each other through background impedance $Z_{\rm b}$.
(b) A coplanar microwave resonator can be modeled as an $LCR$ resonator~\cite{Goppl08}. 
(c) $N$~superconducting transmon artificial atoms~\cite{Koch2007} (crossed boxes) embedded in the coplanar resonator
interact with the cavity mode through coupling capacitors~$C_{{\rm g}i}$. 
}\label{fig:setup}
\end{figure}

The system we consider is shown in Fig.~\ref{fig:setup}(a).
Microwaves propagate in two semi-infinite transmission lines (TLs). The TLs are connected to each other
through a two-sided cavity and through a background. The cavity is
described by a resonant element $Z$ and coupling capacitors $C_{\rm c i}$. 
The background coupling between the TLs is described by a parallel impedance $Z_{\rm b}$.
Multiple superconducting artificial atoms (transmons) can be embedded in the cavity, affecting its resonance frequency,
as described below.

We start building a theoretical model for this circuit in Sec.~\ref{sec:TransmissionLine} by introducing a quantized model of microwave radiation in TLs.
In Sec.~\ref{sec:CouplingToCavity}, we account for the coupling between the cavity and the TLs and in  Sec.~\ref{sec:Quantization}
between the cavity and the artificial atoms. In Sec.~\ref{sec:ParallelHeisenbergEquations}, we show how the parallel transmission channel can be included in the model by additional terms in microwave boundary conditions.
A full solution in the linear limit is derived in Sec.~\ref{sec:LinearSolution} and
a master equation for simulation in a more general situation is given in Sec.~\ref{sec:GeneralSolution}.

\subsection{Microwave radiation in open transmission lines}\label{sec:TransmissionLine}

\subsubsection{Radiation states}
Microwave radiation in the left TL can be described by a traveling-field solution for the magnetic flux~\cite{WallsMilburn,Loudon,Pozar,Hofheinz2018}
\begin{eqnarray}\label{eq:MW1}
&&\hat \Phi(x<0,t)=\\
&&\sqrt{\frac{\hbar Z_{0}}{2\omega_{0}}}\left[ \hat a_{\rm  in}^{\rm L} (t-x/c)+ \hat a_{\rm out}^{\rm L} (t+x/c)+{\rm H.c.} \right]\, .\nonumber
\end{eqnarray}
We associate the position $x=0$ as the cavity boundary and the space is semi-infinite, $-\infty<x<0$.
The capacitance $C'$ and inductance $L'$ per unit length define the characteristic impedance $Z_0=\sqrt{L'/C'}$ and 
the effective speed of light $c=1/\sqrt{L'C'}$.
We work in the narrow-bandwidth approximation~\cite{Loudon} and therefore the cavity resonance frequency $\omega_0$ (defined more detailed below) appears in the equation.
The operator $\hat a_{\rm in}^{\rm L}(t)$  annihilates and $\hat a_{\rm in}^{\rm L}(t)^{\dagger}$ creates an incoming photon.
These operators satisfy the commutation relations
\begin{eqnarray}
\left[\hat a_{\rm in}^{\rm L}(t),\hat a_{\rm in}^{\rm L}(t')^{\dagger} \right]&=&\delta(t-t')\label{eq:InputCommutation1}  \, .
\end{eqnarray}
The same relation is valid also for the corresponding outgoing-field photon operators $\hat a_{\rm out}^{\rm L}(t)$.
We can also define photon operators of specific frequencies
\begin{eqnarray}\label{eq:PC1}
\hat a_{\rm in/out}^{\rm L}(\omega)=\frac{1}{\sqrt{2\pi}}\int_{-\infty}^\infty dt e^{i\omega t} \hat a_{\rm in/out}^{\rm L}(t) \, .
\end{eqnarray}
We have then
\begin{eqnarray}\label{eq:BackToOriginal}
\hat a_{\rm in/out}^{\rm L}(t)=\frac{1}{\sqrt{2\pi}}\int_{-\infty}^\infty d\omega e^{-i\omega t} \hat a_{\rm in/out}^{\rm L}(\omega) \, .
\end{eqnarray}
The commutation relations of these fixed-frequency operators have the form
\begin{equation}\label{eq:CR1}
\left[ \hat a_{\rm in}^{\rm L}(\omega),\hat a_{\rm in}^{\rm L}(\omega')^{\dagger} \right]=\delta(\omega-\omega') \, .
\end{equation}

A similar definition is made for the operators decribing fields on the right of the cavity,
$\hat a_{\rm in}^{\rm R}$ and $\hat a_{\rm out}^{\rm R}$.

\subsubsection{Microwave transmission and reflection}

The microwave properties we study in this paper are the microwave reflection, $s_{11}(\omega)$, and the microwave transmission, $s_{12}(\omega)$.
They are defined through the output amplitudes when having a coherent input from one side and no input from the other side.
By assuming a coherent input of frequency $\omega$ from the left, we define
\begin{eqnarray}
s_{11}(\omega)&=&   \frac{\langle \hat a_{\rm out}^{\rm L}(\omega)\rangle }{\langle \hat a_{\rm in}^{\rm L}(\omega)\rangle} \label{eq:ReflectionOperatorDefinition1} \\
s_{12}(\omega)&=&   \frac{\langle \hat a_{\rm out}^{\rm R}(\omega)\rangle}{\langle \hat a_{\rm in}^{\rm L}(\omega) \rangle} \label{eq:ReflectionOperatorDefinition2} \, .
\end{eqnarray}
In theoretical modeling,
these relations can be determined through solving microwave boundary conditions and cavity equations of motion, as described below.


\subsection{Open transmission line connected to cavity}\label{sec:CouplingToCavity}

We continue building the model by considering first a situation
where there are no transmons embedded in the cavity, i.e., the setup of Fig.~\ref{fig:setup}(b).
We further assume the absence of parallel transmission ($Z_{\rm b}=\infty$) and that the cavity has no intrinsic dissipation ($R=\infty$).
We describe the photonic state inside the cavity by a single-mode Hamiltonian
\begin{eqnarray}
\hat H_0=\hbar\omega_0\hat a^\dagger \hat a \, .
\end{eqnarray}
Here the operator $\hat a^{(\dagger)}$ is the cavity photon annihilation (creation) operator satisfying $[\hat a,\hat a^{\dagger}]=1$.
The coupling-normalized resonance frequency is 
\begin{eqnarray}
\omega_0=\frac{1}{\sqrt{L(C+2C_{\rm c})}} \, .
\end{eqnarray}
Here $L$ ($C$) is the inductance (capacitance) of the cavity and $C_{\rm c}$ the coupling capacitance between the
cavity and a TL. From here on we assume the case of symmetric coupling, $C_{\rm c1}=C_{\rm c2}=C_{\rm c}$.

The interaction between the semi-infinite TLs and the cavity is described by boundary conditions at the two sides of the cavity~\cite{WallsMilburn}. On the left-hand side we write
\begin{eqnarray}
\hat a_{\rm out}^{\rm L}(t)&=& \sqrt{\gamma}\hat a(t)-\hat a_{\rm in}^{\rm L}(t) \, . \label{eq:BoundaryConditionNoParallel1}
\end{eqnarray}
The operators are time dependent since the condition is given in the Heisenberg picture.
Similarly for the right-hand side,
\begin{eqnarray}
\hat a_{\rm out}^{\rm R}(t)&=&\sqrt{\gamma}\hat a(t)-\hat a_{\rm in}^{\rm R}(t) \, . \label{eq:BoundaryConditionNoParallel2} 
\end{eqnarray}
The decay rate is now identical in the two directions and has the form
\begin{eqnarray}
\gamma &=& \left( \frac{C_{\rm c}}{C+2C_{\rm c}} \right)^2 \frac{Z_0}{Z_{LC}} \omega_0 \,,
\end{eqnarray}
where the characteristic impedance of the resonator is $Z_{LC}=\sqrt{L/(C+2C_{\rm c})}$.
This treatment of the cavity field is valid for high quality factors, $Q=\omega_0/2\gamma\gg 1$. 

The cavity field operator also satisfies the Heisenberg equation of motion~\cite{WallsMilburn}
\begin{equation}
\hat{\dot {a}}(t) = \frac{i}{\hbar}\left[\hat H_0,\hat a(t)\right] -\gamma\hat a(t) +\sqrt{\gamma} \left[ \hat a_{\rm in}^{\rm L}(t)+\hat a_{\rm in}^{\rm R}(t) \right]  \,. \label{eq:HeisenbergEquation}
\end{equation}
This is found to be also more generally valid, with proper redefinition of Hamiltonian $\hat H_0$, accounting for the presence of artificial atoms, and also in the presence of reactive and dissipative parallel coupling.

\subsection{Cavity interacting with artificial atoms}\label{sec:Quantization}
It is straightforward to account for the presence of artificial atoms in the above treatment.
We first consider the case of including them as two-level systems and after this
generalize the treatment to the case of multi-level atoms.

\subsubsection{Interaction with two-level systems}\label{sec:DispersiveTwoLevels}
For transverse cavity-atom couplings $g_i\ll \omega_0$, the system is decribed in the rotating-wave approximation by
the Tavis-Cummings Hamiltonian~\cite{WallsMilburn}
\begin{equation}
\hat H_0=\hbar\omega_0\hat a^\dagger \hat a+\hbar\sum_{i=1}^n\frac{\Omega_i}{2}\hat\sigma_z^i+\sum_{i=1}^n\hbar g_i \left(\hat a^\dagger \hat \sigma_-^i + \hat a \hat \sigma_+^i \right) \, .
\end{equation}
Here $\hat\sigma_{+(-)}^i$ is the spin raising (lowering) operator of two-level system~$i$.
Boundary conditions~(\ref{eq:BoundaryConditionNoParallel1}-\ref{eq:BoundaryConditionNoParallel2}) and Heisenberg equation of motion~(\ref{eq:HeisenbergEquation}) keep their form.

Here, when $g_i\ll \vert\Delta_i\vert$, where  $\Delta_i=\Omega_i-\omega_0$,
the system shows effectively a longitudinal coupling between the resonator and the two-level systems.
This is called a dispersive coupling regime.
For a single-atom environment, the resulting Hamiltonian has the form~\cite{Blais2004,Zueco2009}
\begin{align}\label{eq:DispersiveHamiltonian}
\hat H_0'&=\hbar\left( \omega_0 +\chi\hat\sigma_z   \right)\hat a^\dagger\hat a +\frac{\hbar}{2}\left(\Omega+\chi\right)\hat\sigma_z \, .
\end{align}
Here we have defined the dispersive shift
\begin{align}\label{eq:DispersiveShiftSingle}
\chi&=\frac{g^2}{\Delta}\, .
\end{align}
Corrections to this Hamiltonian are higher orders in $g/\Delta\ll 1$. We see that
the effective resonance frequency of the cavity then depends on the state of the two-level system and can have values $\omega_0 \pm \chi$.
This type of Hamiltonian can be used to describe a driven cavity if the neglected energy-level anharmonicity
is small compared to the energy-level broadening~\cite{Bishop2010}. For low photon-number distributions the relevant condition is $g^2/\vert\Delta\vert \times (g/\Delta)^2\ll \gamma$, which is assumed to be true when considering a dispersive regime in this article.
This coupling regime is commonly used for readout of superconducting qubits~\cite{Wallraff2005}.

A generalization of the dispersive Hamiltonian to account for multiple two-level systems is straightforward,
\begin{align}\label{eq:DispersiveHamiltonian2}
\hat H_0'&=\hbar\left( \omega_0 +\sum_{i=1}^n\chi_i\hat\sigma_z^i   \right)\hat a^\dagger\hat a +\frac{\hbar}{2}\sum_{i=1}^n\left(\Omega_i+\chi_i\right)\hat\sigma_z^i \, . \nonumber
\end{align}
Here $\chi_i=g_i^2/\Delta_i$, $g_i\ll \vert\Delta_i\vert$, and we have neglected cavity-mediated couplings
between the two-level systems~\cite{Majer2007,Zueco2009}, i.e., we assume $g_ig_j/\vert \Delta_{i/j}\vert \ll (\Omega_i-\Omega_j)^2$.

\subsubsection{Interaction with multi-level systems}\label{sec:DispersiveManyLevels}

In our experiment, the cavity interacts with transmon artificial atoms.
For multi-level artificial atoms with low anharmonicity like the transmon also higher energy levels need to be accounted for~\cite{Koch2007}. This can be the case even though only two lowest (photon-dressed) atom states would be populated.

A Hamiltonian of a resonator connected to one multi-level atom can be written generally in the form
\begin{align}
\hat H_0&= \hbar\omega_{0} \hat a^\dagger \hat a+\sum_j E_j \vert j\rangle\langle j\vert +\sum_{k,l}\hbar g_{kl}(\hat a+\hat a^\dagger)\vert k\rangle\langle l\vert \, .
\end{align}
Here $E_j$ is the energy of the atom-state $\vert j\rangle$ and $j=0,1,2,\ldots$.
The effective Hamiltonian in the dispersive regime has here the form
\begin{align}
\hat H_0'&=\hat H_{\rm C}+ \sum_j E_j \vert j\rangle\langle j\vert +\sum_{j>0}\hbar\chi_{j-1,j}\vert j\rangle\langle j\vert \, ,
\end{align}
where the resonator-atom coupling is described by
\begin{align}
\hat H_{\rm C}=\hbar \hat a^\dagger \hat a \left( \omega_{0}-\chi_{01}\vert 0\rangle\langle 0\vert +\sum_{j>0}(\chi_{j-1,j}-\chi_{j,j+1})\vert j\rangle\langle j\vert  \right) \, .
\end{align}
and
\begin{align}
\chi_{j-1,j}&=j\frac{g^2}{E_{j}/\hbar-E_{j-1}/\hbar-\omega_{0}} \, .
\end{align}

If only two lowest (photon-dressed) transmon levels are populated,
the correction to the dispersive-shift operator of Eq.~(\ref{eq:DispersiveHamiltonian}) is accounted for by the replacement~\cite{Koch2007}
\begin{eqnarray}\label{eq:DispersiveHamiltonianTransmon}
\frac{g^2}{\Delta} \hat\sigma_z  \hat a^\dagger\hat a \leftarrow \left(\frac{g^2}{\Delta}-\frac{g^2}{\Delta-E_C/\hbar} \right) \hat\sigma_z  \hat a^\dagger\hat a \, .
\end{eqnarray}
Here the charging energy $E_C$ corresponds to the anharmonicity of the artificial atom.

Finally, transmons can be described as parallel $L_iC_i$ 
circuits with coupling capacitors $C_{{\rm g}i}$, see Fig.~\ref{fig:setup}(c).
The explicit form of the cavity-qubit couplings can then be expressed as a function of
these linear-circuit elements~\cite{Koch2007}. They have here the form
\begin{eqnarray}
\hbar g_i= \frac{\hbar}{2}\sqrt{\omega_0\Omega_i} \frac{C_{{\rm g}i}}{\sqrt{C C_i}}   \, .
\end{eqnarray}
We assume here small coupling capacitors, $C_{{\rm g}i},C_{\rm c}\ll C,C_{i}$.
The lowest energy-level splittings are in the same limit $\Omega_{i}= 1/\sqrt{L_i(C_i+C_{{\rm g}i})}$, the pure cavity frequency $\omega_0= 1/\sqrt{ L(C+2C_{\rm c}+\sum_iC_{{\rm g}i})}$, and 
the charging energies $E_C^i= e^2/2(C_i+C_{{\rm g}i})$.

The dispersive shifts of a multi-transmon system as well as of higher excited states of single transmons have been studied experimentally in Refs.~\cite{Reed2010,Peterer2015,Braumuller2015}.

\begin{figure}
\includegraphics[width=\columnwidth]{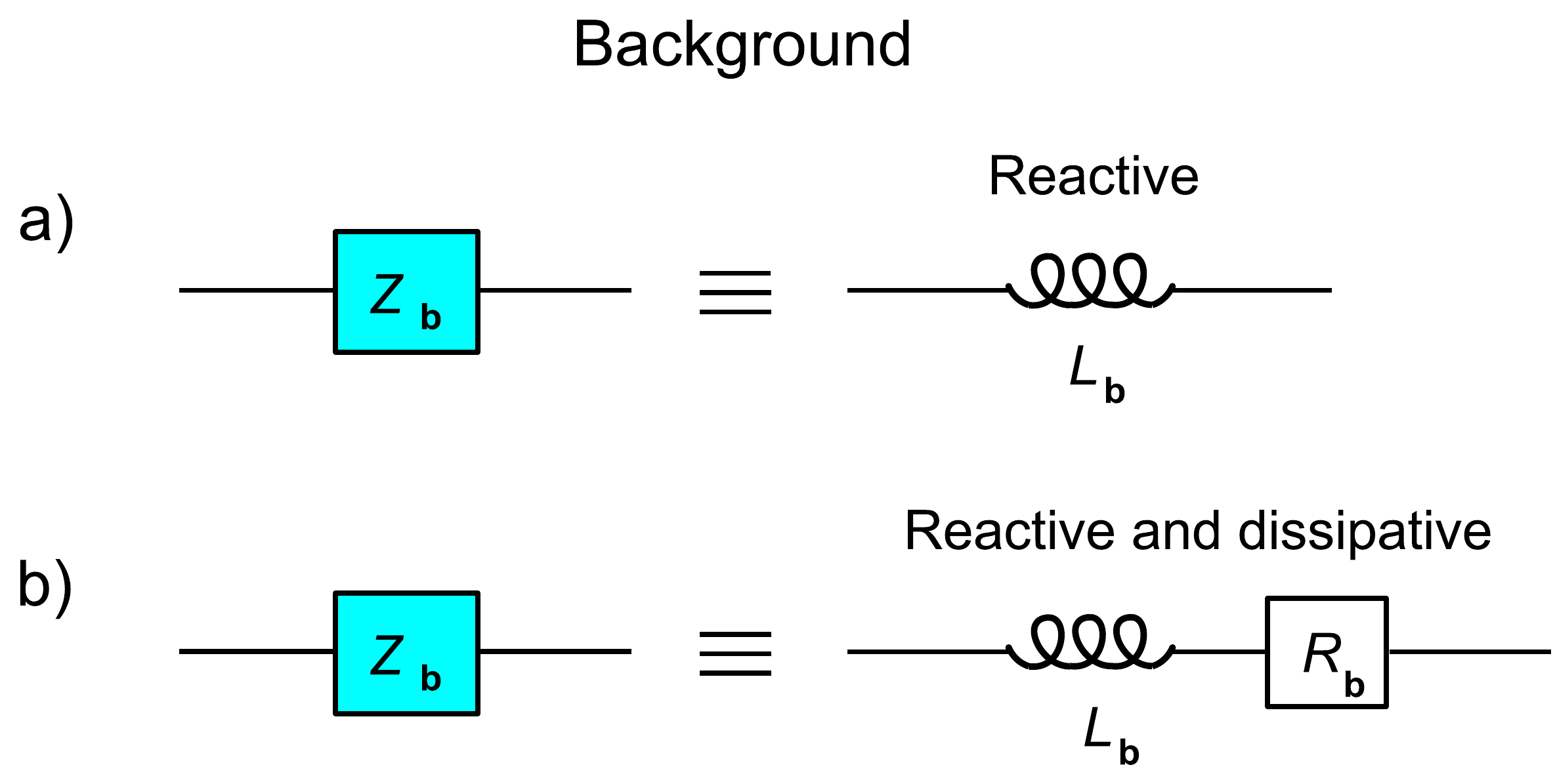}
\caption{
(a) A background described by an inductor $L$ connecting the input and output transmission lines.
(b) A background with inductive and dissipative properties, the latter described by a resistor $R_{\rm b}$.
}\label{fig:setup2}
\end{figure}

\subsection{Background}\label{sec:ParallelHeisenbergEquations}
We now extend the above model to also account for a parallel transmission with the cavity, i.e.,
element $Z_{\rm b}$ in Fig.~\ref{fig:setup}(a).
We first consider a reactive (capacitive or inductive) coupling beside the cavity
and after this extend the analysis to also account for dissipative background.

\subsubsection{Reactive background}
In the case of purely reactive background, as in Fig.~\ref{fig:setup2}(a),
the  boundary conditions of Eqs.~(\ref{eq:BoundaryConditionNoParallel1}-\ref{eq:BoundaryConditionNoParallel2})
can be shown to generalize to (Appendix~B)
\begin{align}
\hat a_{\rm out}^{\rm L}(t) &=  \sqrt{\gamma} \hat a(t)- \frac{1}{1+2i\epsilon}\hat a_{\rm in}^{\rm L}(t)-\frac{2i\epsilon}{1+2i\epsilon}\hat a_{\rm in}^{\rm R}(t)\label{eq:CavityFanoOuta} \\
\hat a_{\rm out}^{\rm R}(t) &=  \sqrt{\gamma} \hat a(t)- \frac{1}{1+2i\epsilon}\hat a_{\rm in}^{\rm R}(t)-\frac{2i\epsilon}{1+2i\epsilon}\hat a_{\rm in}^{\rm L}(t)\label{eq:CavityFanoOutb} \,.
\end{align}
Here we have introduced a parameter describing the reactive response of a parallel inductance $L_{\rm b}$
\begin{eqnarray}\label{eq:EpsilonParallelInductance}
\epsilon&=&\frac{Z_0}{\omega_0L_{\rm b}}=\frac{Z_0}{\vert Z_{\rm b}(\omega_0)\vert} \,.
\end{eqnarray}
We consider explicitly the case of a parallel inductor, whereas the result for a parallel capacitor is obtained by a sign change
$\epsilon\rightarrow -\epsilon$ (Appendix~C). 
The limit $\epsilon\rightarrow 0$ then gives the previous input-output relations for two-sided cavity, Eqs.~(\ref{eq:BoundaryConditionNoParallel1}-\ref{eq:BoundaryConditionNoParallel2}).
Essential is that the Heisenberg equation of motion for the cavity field, Eq.~(\ref{eq:HeisenbergEquation}), stays the same, 
irrespective of the values of $\gamma $ and $\epsilon$.

It should be emphasized that since
the Heisenberg equation of motion for the cavity field stays the same,
models of driven cavities~\cite{Bishop2009,Bishop2010,Reed2010,Fink2017} or dispersive
quantum-state measurement~\cite{Wallraff2005,Koch2007}
can be directly generalized to cover also the presence of background transmission, just by using the modified boundary conditions~(\ref{eq:CavityFanoOuta}-\ref{eq:CavityFanoOutb})
when evaluating properties of the outgoing fields (from the unchanged solution for the cavity field).

\subsubsection{Dissipative background}
A finite resistivity of the background can be introduced by adding a negative imaginary part to parameter $\epsilon$, i.e., by replacing
\begin{eqnarray}\label{eq:DissipativeBackground}
\epsilon\leftarrow  \epsilon-i\epsilon_{\rm d} \, .
\end{eqnarray}
The exact relation to the model shown in Fig.~\ref{fig:setup2} can be derived to be (Appendix~A)
\begin{align}
\epsilon-i\epsilon_{\rm d}&=-i\frac{Z_0}{Z^*_{\rm b}(\omega_0)}=\frac{Z_0(\omega_0 L_{\rm b}-iR_{\rm b})}{\omega_0^2 L_{\rm b}^2+R_{\rm b}^2  } \\
Z_{\rm b}(\omega)&=i\omega L_{\rm b}\omega+R_{\rm b} \, .
\end{align}
The model is valid for arbitrary strengths of the background transmission, as long as the cavity mode is coupled only weakly to the transmission lines and the background changes weakly within cavity linewidth $\gamma$.
Note that variables $\epsilon$ and $\epsilon_{\rm d}$ are dimensionless.

\subsubsection{Removing the background from experimental data}

We can also remove the Fano resonance from an experimental data to recover the pure cavity spectrum.
For example, 
if we have only input from one side of the cavity and we  have measured the
transmission also when artificial atoms and the resonator are tuned far away, which result we mark now
$s_{12}^{\rm background}=2(\epsilon-i\epsilon_{\rm d})/[{\rm i}-2(\epsilon-i\epsilon_{\rm d})]$,
the above results imply that
\begin{align}
\frac{\sqrt{\gamma}}{\left\langle \hat a_{\rm in}^{\rm L} \right\rangle}  \left\langle \hat a \right\rangle &=  \left(  s_{12}  -s_{12}^{\rm background}\right)  \label{eq:Recovery1} \,.
\end{align}
Here $s_{12}$ is the measured transmission in the presence of resonator and artificial atoms.
Since the cavity equation of motion is independent of the background, the left-hand side of Eq.~(\ref{eq:Recovery1})
is also the transmission in the abence of the background. We must have then
\begin{align}
 s_{12}^{\rm free} &=   s_{12}  -s_{12}^{\rm background}  \label{eq:Recovery2} \,.
\end{align}
Here $s_{12}^{\rm free}$ is the transmission in a hypothetical experiment, where the background is not present.
This relation remains to be valid for all stregths of the background transmission. 

The transformation from the measured background transmission to the cavity boundary parameter $\epsilon$ has the form
\begin{align}
\epsilon-i\epsilon_{\rm d} &= \frac{i}{2} \frac{s_{12}^{\rm background}}{1+s_{12}^{\rm background}}   \label{eq:EpsilonInverse} \,.
\end{align}

\subsection{Solution for a linear cavity}\label{sec:LinearSolution}

In the case of a linear cavity, we can solve the out-fields as a function of input directly by Fourier transformation.
This solution is valid also for a cavity-atom system in the dispersive limit when no atom transitions occur, or when transitions are slow and can be accounted for by statistical averaging.

In the case of a dissipationless cavity and background we have
\begin{eqnarray}\label{eq:SolutionSeriesCapacitor}
\left(
\begin{matrix}
\hat a_{\rm out}^{\rm L}(\omega) \\
\hat a_{\rm out}^{\rm R}(\omega)
\end{matrix}
\right)
&=&\frac{1}{(1+2i\epsilon)(1-2if)}\\
&\times&\left(
\begin{matrix}
4\epsilon f -1 &  - 2i(\epsilon + f) \\
- 2i(\epsilon + f) &  4\epsilon f -1
\end{matrix}
\right)
\left(
\begin{matrix}
\hat a_{\rm in}^{\rm L}(\omega) \\
\hat a_{\rm in}^{\rm R}(\omega)
\end{matrix}
\right) \, , \nonumber
\end{eqnarray}
where
\begin{eqnarray}
f(\omega)=\frac{\gamma}{2(\omega_0-\omega)} \, .
\end{eqnarray}
The possible dispersive shift of the cavity frequency is now incorporated in $\omega_0$.
Here we consider explicitly the case of a non-dissipative background ($\epsilon_{\rm d}=0$),
but the result in the general case can be obtained by replacement~(\ref{eq:DissipativeBackground}).

The scattering amplitude can be shown to be here
\begin{eqnarray}\label{eq:SolutionSeriesCapacitorAmplitude}
s_{12}&=& \frac{2\epsilon}{i-2\epsilon}+\frac{\gamma}{\gamma+i(\omega_0-\omega)} \\
   &=& \frac{ \gamma +2\epsilon(\omega_0-\omega)     }{(1+2i\epsilon)\left[\gamma+i(\omega_0-\omega)  \right]}  \,. \nonumber
\end{eqnarray}
The special cases $\epsilon=0$ or $\gamma=0$ give the transmission amplitudes when the parallel transmission does not contribute
or the cavity does not contribute, correspondingly.
The reflection amplitude has the form
\begin{eqnarray}\label{eq:SolutionSeriesCapacitorAmplitudeReflection}
s_{11}&=& -\frac{1}{1+2i\epsilon}+\frac{\gamma}{\gamma+i(\omega_0-\omega)} \\
   &=& \frac{ 2\epsilon\gamma +\omega-\omega_0   }{(-i+2\epsilon)\left[\gamma+i(\omega_0-\omega)  \right]}  \,. \nonumber
\end{eqnarray}
For a non-dissipative system the transmission and reflection powers sum to $1$,
\begin{eqnarray}
\vert s_{11}\vert^2+\vert s_{12}\vert^2=1 \,.
\end{eqnarray}
Here, the solution also satisfies the commutation relations
\begin{eqnarray}
\left[\hat a_{\rm out}^{\rm L}(t),\hat a_{\rm out}^{\rm L}(t')^\dagger \right]&=&\delta(t-t')  \\
\left[\hat a_{\rm out}^{\rm R}(t),\hat a_{\rm out}^{\rm R}(t')^\dagger \right]&=&\delta(t-t')  \, ,
\end{eqnarray}
which is obtained only by assuming that this is true for the input fields, Eq.~(\ref{eq:InputCommutation1}), demonstrating consistency of the theory.

In the linear solution, the effect of intrinsic dissipation of the cavity field can be accounted for
by adding an imaginary part to the resonance frequency, i.e., replacing
\begin{eqnarray}
\omega_0\leftarrow  \omega_0-i\frac{\kappa}{2} \, .
\end{eqnarray}
This corresponds in the parallel $LCR$-circuit of Fig.~\ref{fig:setup} to~\cite{Pozar}
\begin{align}
\kappa=\frac{1}{RC}=\frac{\omega_0}{Q_{\rm int}} \, ,
\end{align}
where in the second form we have defined an internal quality factor $Q_{\rm int}=\omega_0RC$.
The equivalent energy decay rate is then
\begin{eqnarray}
\frac{1}{RC}=\kappa \, .
\end{eqnarray}
This approach assumes implicitly $\kappa\ll \omega_0$.

At finite temperatures, a dissipative system performs fluctuations. In our analysis, based on
expectation values of Eqs.~(\ref{eq:ReflectionOperatorDefinition1}-\ref{eq:ReflectionOperatorDefinition2}),
temperature plays a role only in the case of a non-linear system, since otherwise thermal fluctuations average out.
This is since for a linear system, thermal fluctuations result in an additional width of the Gaussian probability distribution of the field quadratures around the classical mean~\cite{WallsMilburn}. Similarly, thermal radiation emitted by the background
is not expected to contribute to average transmission and reflection.


\subsection{General system: Master equation simulation}\label{sec:GeneralSolution}
We can simulate a more general cavity-atom system using a Lindblad master equation~\cite{WallsMilburn}
\begin{equation}\label{eq:MasterEquation}
\hat{\dot {\rho}}=\frac{i}{\hbar}[\hat \rho,\hat H]+{\cal L}_{\rm L}[\hat \rho]+{\cal L}_{\rm R}[\hat \rho]+\sum_{i=1}^n{\cal L}_{{\rm q}}^i[\hat \rho] \, .
\end{equation}
Here $\hat \rho$ is the reduced density matrix of the resonator and artificial atoms.
The Hamiltonian $\hat H$ now accounts for a coherent drive, i.e., operator $\hat a_{\rm in}^{\rm L}$, in a way shown below.
Lindblad operators ${\cal L}_{{\rm L/R}}$ then describe decay to (and excitations from) TLs and Lindblad operators ${\cal L}^{i}_{\rm q}$ decoherence (decay and dephasing) of artificial atoms.

\subsubsection{Lindblad operators}\label{sec:LindbladOperators}
As previously, we assume that cavity-qubit couplings are small, $g_i\ll\omega_0,\Omega_i$, and that possible differences between decay rates related to different shifted values of the cavity frequency can be neglected.
The Lindblad super-operator ${\cal L}_{\rm L}$ then describes cavity photon transitions due to interaction with the left TL,
\begin{eqnarray}
{\cal L}_{\rm L}[\hat \rho]&=&\frac{\gamma^-}{2}\left( 2\hat a \rho \hat a^\dagger -\hat a^\dagger \hat a \rho-\rho\hat a^\dagger \hat a \right)\nonumber \\
&+&\frac{\gamma^+}{2}\left( 2\hat a^\dagger \rho \hat a - \hat a \hat a^\dagger \rho-\rho \hat a  \hat a^\dagger   \right)\, .
\end{eqnarray}
The decay rate to the left TL satisfies in thermal equilibrium
\begin{eqnarray}\label{eq:ThermalEquilibrium1}
\gamma^{-}=\gamma\left[ 1+\frac{1}{\exp\left(\frac{\hbar\omega_0}{k_{\rm B}T}\right)-1}  \right]  \, ,
\end{eqnarray}
and correspondingly for the thermal excitation rate
\begin{eqnarray}\label{eq:ThermalEquilibrium2}
\gamma^{+}=\gamma\left[ \frac{1}{\exp\left(\frac{\hbar\omega_0}{k_{\rm B}T}\right)-1}  \right]  \, .
\end{eqnarray}
We have then $\gamma^{-}=\gamma^{-}(T=0)+\gamma^{+}$.
Similarly for the interaction with right TL described by super-operator ${\cal L}_{\rm R}$.

Intrinsic dissipation and fluctuations of artificial atoms can also be added by Lindblad super-operators.
In the case of two-lvel systems we have
\begin{eqnarray}
{\cal L}_{{\rm q}}^i[\hat \rho]&=&\frac{\kappa_{i}^-}{2}\left( 2\hat \sigma_-^i \rho \hat \sigma_+^i -\hat \sigma_+^i \hat \sigma_-^i \rho-\rho\hat \sigma_+^i \hat \sigma_-^i \right)\nonumber\\
&+&\frac{\kappa_{i}^+}{2}\left( 2\hat \sigma_+^i \rho \hat \sigma_-^i -\hat \sigma_-^i \hat \sigma_+^i \rho-\rho\hat \sigma_-^i \hat \sigma_+^i \right)\, ,
\end{eqnarray}
where $\kappa_{i}^\pm$ are the corresponding transition rates of two-level system~$i$.
These rates are affected by the electromagnetic environment as seen by the transmons~\cite{Leppakangas2018}.
The temperature dependence of $\kappa^{\pm}_i$ is equivalent to Eqs.~(\ref{eq:ThermalEquilibrium1}-\ref{eq:ThermalEquilibrium2}). 

Additionally, qubit pure dephasing~\cite{Ithier2005}  can be accounted for within an operator
\begin{eqnarray}\label{eq:PureDephasing}
{\cal L}_{\phi}[\hat \rho]&=&\sum_i\frac{\kappa_{\phi}^i}{2}\left( \hat \sigma_z^i \rho \hat \sigma_z^i - \rho\right) \, .
\end{eqnarray}
Similarly, it is possible to account for cavity dephasing by
replacement $\hat\sigma_z\leftarrow \hat \sigma_a=2\hat a^\dagger \hat a-1$.
In the experiment analysis, we find that 
resonator dephasing due to coupling to multiple atoms can be modeled rather well within such model.
Furthermore, in the linear case (Sec.~\ref{sec:LinearSolution}), such cavity dephasing can also be accounted for as
cavity loss ($\kappa>0$), discussed more detailed below in Sec.~\ref{sec:Dephasing}.

\subsubsection{Coherent drive in the Hamiltonian}
We consider the case of an incoming coherent radiation from the left-hand side TL.
The presence of such coherent drive is accounted for by an effective Hamiltonian~\cite{WallsMilburn}
\begin{eqnarray}\label{eq:DispersiveDrive}
\hat H= \hat H_{0}+\hat H_{\rm d} \, ,
\end{eqnarray}
where the drive appears as a term
\begin{eqnarray}\label{eq:NewDriveHamiltonian}
\hat H_{\rm d}=   i\hbar\sqrt{\gamma}A (t) \hat a^\dagger  +  {\rm H.c.} \, .
\end{eqnarray}
This form is derived by assuming
\begin{eqnarray}
\left\langle \hat a_{\rm in}^{\rm L}(t)   \right\rangle=A(t) \, ,
\end{eqnarray}
and $\left\langle \hat a_{\rm in}^{\rm R}(t)   \right\rangle=0$. 
For example, in the dispersive regime,
the corresponding total Hamiltonian of a driven cavity coupled to single two-level system can be written in the form
\begin{align}
\hat H_0'&=\hbar\left( \omega_0-\omega +\chi\hat\sigma_z   \right)\hat a^\dagger\hat a +\frac{\hbar}{2}\left(\Omega_0+\chi\right)\hat\sigma_z \nonumber  \\
&+\frac{\alpha}{2}\left(\hat a +\hat a^\dagger\right) \,.
\end{align}
Here we have gone into the rotating frame with respect to drive frequency $\omega$ and assumed that $\alpha=2i\hbar\sqrt{\gamma}A(t)e^{i\omega t}$ is a real number (so that $A(t)\propto i e^{-i\omega t}$).
The generalization to multi-level atoms and multi-atom systems is straightforward.

\subsubsection{Solution for the out field}\label{sec:SolutionMethod}
For obtaining the average output fields, $ \langle \hat a_{\rm out}(t)  \rangle$ and $ \langle \hat b_{\rm out}(t)   \rangle$,
we first determine the steady-state solution for the cavity field, $\left\langle \hat a(t)   \right\rangle$, obtained from
solving the master equation~(\ref{eq:MasterEquation}) with Hamiltonian~(\ref{eq:DispersiveDrive}). 
After this the solution for the out-field  is obtained from the boundary conditions~(\ref{eq:CavityFanoOuta}-\ref{eq:CavityFanoOutb}) with
inserting the assumed form of the input field $ \left\langle \hat a_{\rm in}^{\rm L}(t)   \right\rangle=A(t)$.


\section{Fano resonances and the effect of decoherence}\label{sec:FanoResonance}

In this section, we investigate microwave transmission across a two-sided cavity in the presence of background
transmission and decoherence (dissipation and fluctuations).
We start in Sec.~\ref{sec:FanoGeneral} by giving a short summary of the form of the conventional Fano-resonance.
In Sec.~\ref{sec:DissipationFreeFano} we solve the Fano resonance in the case of a linear cavity
with no internal or background dissipation.
In Sec.~\ref{sec:DissipativeFano}, we study the influence of dissipation in the case of a linear cavity.
In Sec.~\ref{sec:JaynesCummingsOscillatorDecoherence}, we study 
a cavity connected to single or multiple dissipative two-level systems subjected to heating and fluctuations.
Finally, in Sec.~\ref{sec:HigherEnergyLevels}, we estimate the size of contribution from higher energy levels
of transmons.





\subsection{Conventional Fano function}\label{sec:FanoGeneral}

In a Fano resonance~\cite{Fano1961,FanoNanoStructures2010,FanoPhotonics2017}, the spectral response of a resonant system is asymmetric around the resonance frequency due to an interference effect between two scattering amplitudes: scattering through a background with a constant (or wide) state density and scattering through a  discrete (or narrow) energy-level. 
The conventional form of the Fano interference is characterized by only single variable: Fano parameter $q$.
Here, the total scattering amplitude $\vert s\vert$, or spectral density $\vert s\vert^2$, are of the form (neglecting normalization factors)
\begin{eqnarray}\label{eq:FanoGeneral}
\vert s\vert \propto \frac{\vert q+\eta\vert}{\sqrt{1+\eta^2}}\,\,\, , \,\,\, \vert s\vert^2 \propto \frac{( q+\eta)^2}{1+\eta^2}\, ,
\end{eqnarray}
where $\eta$ is a broadening-normalized drive frequency with respect to the resonance frequency,
$\eta=(\omega_0-\omega)/(\Gamma/2)$, with $\Gamma$ being a parameter describing the resonant-state broadening.
Two central limits of this function are $q\rightarrow\infty$, giving a Lorentzian shaped {\em peak}, and $q=0$, giving a a Lorentzian shaped {\em dip}.
This description then catches resonant enhancement as well as resonant suppression as two limits of one formalism.

\subsection{Decoherence-free linear oscillator}\label{sec:DissipationFreeFano}

Consider first the case of dissipation-free linear cavity. 
Here one can use the analytical solutions of Eqs.~(\ref{eq:SolutionSeriesCapacitorAmplitude}-\ref{eq:SolutionSeriesCapacitorAmplitudeReflection}), which give
\begin{eqnarray}
\vert s_{12}\vert^2 = \frac{1}{1+q^2} \frac{(q+\eta)^2}{1 + \eta^2    }  \, .
\end{eqnarray}
Here we have identified
\begin{eqnarray}
q&=&\frac{1}{2\epsilon} \\
\eta&=& \frac{\omega_0-\omega}{\gamma} \, .
\end{eqnarray}
From Eq.~(\ref{eq:EpsilonParallelInductance}) we obtain that for $\vert Z_{\rm b}(\omega_0)\vert\rightarrow\infty$, $q\rightarrow\infty$, and transmission probability $\vert s_{12}\vert^2$ is a Lorentzian peak.
For any finite parallel coupling (finite $q$) interference occurs, which
is perfectly destructive when $q=-\eta$, meaning 
\begin{eqnarray}
\omega=\omega_0+\frac{\gamma}{2\epsilon} \, .
\end{eqnarray}
The response near the resonance frequency can also be a dip, $q=0$, when $\vert Z_{\rm b}(\omega_0)\vert\rightarrow 0$.
Such form then needs a very strong parallel transmission.
In the following, we show that
when the system is dissipative, a dip in the transmission can appear also for a weak parallel transmission.


\subsection{Lossy linear oscillator}\label{sec:DissipativeFano}

\begin{figure}
\includegraphics[width=\columnwidth]{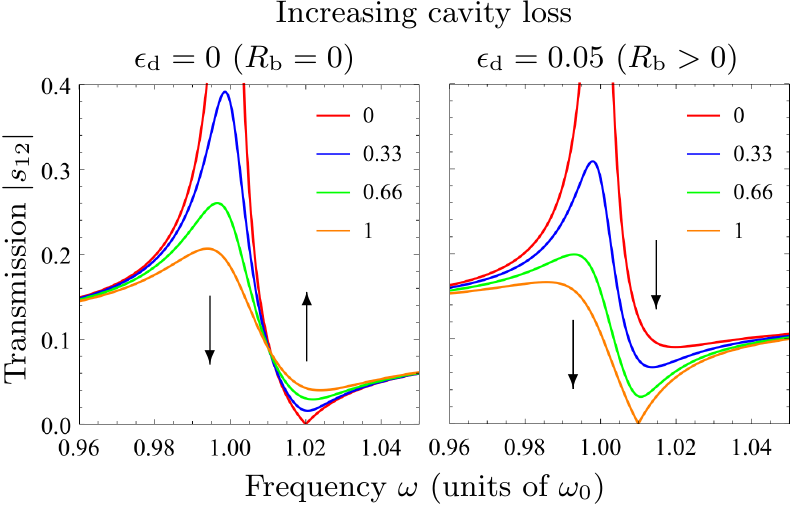}
\caption{ The effect of intrinsic cavity loss on transmission $\vert s_{12}(\omega)\vert$ in the presence of weak background transmission $\epsilon=0.05$.
We increase the dephasing rate $\kappa/2$ as indicated by the plot legends in units of $ 10^{-2}\omega_0$.
When the rate is increased the curve minimum and maximum change as visualized by the arrows. In particular,
without dissipation in the background ($\epsilon_{\rm d}=0$), increase of cavity decoherence lowers the maximum and increases the minimum of $\vert s_{12}(\omega)\vert$.
With a dissipative background  ($\epsilon_{\rm d}=0.05$) both minimum and maximum decrease with increasing cavity decoherence in the considered range $\kappa/2\in[0,0.01]\omega_0$. For higher $\kappa$ the minimum starts to rise.
We consider a linear cavity with resonance frequency $\omega_0$ and coupling to each transmission line  $\gamma=2\times 10^{-3}\omega_0$.
}\label{fig:Theory_Heisenberg_Solution1}
\end{figure}

Consider now including dissipation in the cavity when having a weak parallel transmission, $\epsilon\ll 1$.
A finite intrinsic quality factor of the cavity can be accounted for by adding an imaginary part $\omega_0\leftarrow \omega_0-i\kappa/2$.
The solution of Eq.~(\ref{eq:SolutionSeriesCapacitorAmplitude}) is valid also here.
Transmission $\vert s_{12}(\omega)\vert$ for several values of loss rate $\kappa$ (or equivalently dephasing rate $\kappa/2$) is shown in Fig.~\ref{fig:Theory_Heisenberg_Solution1} ($\epsilon_{\rm d}=0$).
We find that the intrinsic loss reduces transmission and 'straightens' the interference structure.
We note that simultaneously the reflection dip also gets less deep (not plotted).

Consider then adding small dissipation also in the (weak) parallel transmission.
A finite resistivity in the parallel channel can be accounted for by a replacement $\epsilon\leftarrow  \epsilon-i\epsilon_{\rm d}$.
We first note that
without a reactive part in the parallel channel (finite $\epsilon$), we never get an asymmetric resonance curve (tilt). Again,
resistivity in the parallel channel reduces transmission on resonance,
but the two dissipative effects do not simply sum up as an effective increased cavity dissipation rate. 
Instead, if we assume a fixed background dissipation and increase the cavity dissipation (or dephasing), an interesting effect appears:
the minimum value of $\vert s_{12}\vert$ decreases with increasing cavity decoherence and reaches zero, see Fig.~\ref{fig:Theory_Heisenberg_Solution1} ($\epsilon_{\rm d}=0.05$).
Using Eq.~(\ref{eq:SolutionSeriesCapacitorAmplitude}) one can derive that the minimum transmission is exactly zero when 
\begin{eqnarray}\label{eq:DisappearanceCondition}
 \kappa=\gamma\frac{\epsilon_{\rm d}}{\epsilon^2+\epsilon_{\rm d}^2} \,.
\end{eqnarray}
This zero transmission occurs for 
\begin{eqnarray}
\omega= \omega_0+ \frac{\gamma}{2}\frac{\epsilon }{\epsilon^2+\epsilon_{\rm d}^2}   \, . 
\end{eqnarray}

For larger cavity decoherence rates the minimum increases again (plotted later in Fig.~\ref{fig:Comparison11}).
The reflection $\vert s_{11}\vert$, Eq.~(\ref{eq:SolutionSeriesCapacitorAmplitudeReflection}), is also here always a dip.
We then find that an inverted line shape can occur also in the case of weak parallel transmission, when both the cavity and
the background dissipate radiation.

\subsection{Coupling to two-level systems}\label{sec:JaynesCummingsOscillatorDecoherence}
We consider now the case of a cavity with no internal losses ($\kappa=0$) but which
is coupled to a two-level system exhibiting incoherent transitions between its two states.
In particular, we demonstarte two other ways to obtain peak inversion:
(1) fast-incoherent-transitions induced cavity dephasing and (2) reduced resonant-state population. 
Furthermore,
a simple fitting formula combining these two effects is useful in the analysis of experimental data,
as discussed in Sec.~\ref{sec:Experiment}.


\subsubsection{Cavity dephasing due to cavity frequency switching}\label{sec:Dephasing}

\begin{figure}
\includegraphics[width=\columnwidth]{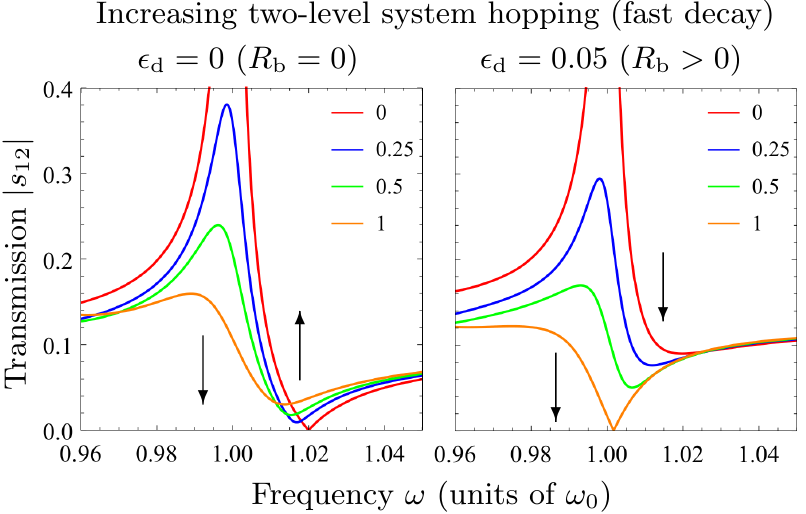}
\caption{
The effect of incoherent transitions in a dispersively coupled two-level system on cavity transmission $\vert s_{12}(\omega)\vert$ in the presence of weak background transmission $\epsilon=0.05$.
The transition rates $\kappa^-$ and  $\kappa^+$ are increased as indicated by the plot legends in units of $ 10^{-2}\omega_0$.
When the transition rates are increased the curve minimum and maximum change as visualized by the arrows.
We consider the case of a large dispersive shift, $g^2/\Delta=-15\gamma$,
and fast decay to ground state, $\kappa^-(T=0)=4 \gamma$. 
The resonance frequency $\omega_0$ accounts for the dispersive shift of the ground state and coupling to each transmission line $\gamma= 2\times 10^{-3}\omega_0$.
}\label{fig:Theory_Stark_Shift_Hopping}
\end{figure}

Stochastic switchings of the two-level systems induce dephasing of the signal propagating through the cavity.
In the dispersive coupling regime, this noise corresponds to stochastic jumping of the frequency between two values,
i.e., to telegraphic noise.
An example of this effect (not mixed with the reduced population effect, Sec.~\ref{sec:ReducedPopulation}) is when one has a high asymmetry between  excitation and decay rates, $\kappa^+\ll \kappa^-$.  
Here,  the two-level system is mostly in its ground state with stochastic short-time visits in its excited state.


Numerical results for the effect of incoherent transitions in a dispersively coupled two-level system on transmission $\vert s_{12}(\omega)\vert$ are shown in Fig.~\ref{fig:Theory_Stark_Shift_Hopping}.
We simultaneously increase the excitation and relaxation rate of the two-level system with the same amount, i.e.,
keeping detailed balance (thermal equilibrium statistics), Eqs.~(\ref{eq:ThermalEquilibrium1}-\ref{eq:ThermalEquilibrium2}).
The total dispersive shift is assumed to be much larger than the cavity broadening, $\vert 2g^2/\Delta\vert \gg \gamma$, and we consider a weak drive power. We find that here incoherent hopping
can create a very similar effect to $s_{12}$ as cavity dissipation, see Fig.~\ref{fig:Theory_Heisenberg_Solution1}.
We however note that for ground-state populations $p<1$, the above dephasing mechanism 
can be differentiated from pure cavity loss by seeking for weak additional peaks (or dips) corresponding dispersive shifts
from the excited states of the artificial atoms.



The similarity to internal decay can be understood as that in this limit a single jump of the two-level system is enough to dephase the system, and that for superpositions of photon numbers the effect of dephasing and decay is qualitatively the same.
More detailed, intrinsic cavity loss with rate $\kappa$ and pure dephasing of the cavity with $\kappa_{\phi}$ affect the average transmission amplitude $s_{12}$ equivalently when
\begin{eqnarray}\label{eq:DephasingConnection}
\frac{\kappa}{2}=\kappa_{\phi} \, .
\end{eqnarray}


In the presence of multiple two-level systems,
the dynamics also have few simple limits.
For relaxation rates much larger than the excitation rates, $\kappa_{i}^- \gg\kappa^{+}_{i}$, with strong dispersive shifts,
the effective dephasing rate is roughly the sum of individual two-level system excitations rates,
\begin{eqnarray}\label{eq:DephasingConnection2}
\kappa_{\phi}\approx \sum_i \kappa^{+}_{i}  \, .
\end{eqnarray}
Furthermore,
the noise spectral density of an ensemble of dispersively-coupled two-level systems is in thermal equilibrium~\cite{Shnirman2005}
\begin{align}
S(\omega)&=\int dt\left\{ e^{i\omega t}\left\langle \hat X(t)\hat X(0) \right\rangle - \left\langle \hat X^2 \right\rangle \right\} \\
&= \sum_i \chi_i^2 \left[ \frac{1}{\cosh^2\left( \frac{\hbar\Omega_i}{2k_{\rm B}T} \right)}  \right] \frac{2\kappa_i}{\omega^2+\kappa_i^2}  \, , \nonumber
\end{align}
where $X=\sum_i\chi_i\sigma_z^i$ and $\kappa_i$ is the decay rate of two-level system $i$ at zero temperature.
This describes random telegraphic noise of thermally excited two-level systems.
For low temperatures we have
\begin{align}
S(\omega)&\approx \sum_i \chi_i^2 \left[ \frac{4}{1+\exp \left(\frac{\hbar\Omega_i}{k_{\rm B}T} \right)}  \right] \frac{2\kappa_i}{\omega^2+\kappa_i^2}  \, ,\\
&= 8\sum_i \left(\frac{\chi_i}{\kappa_i}\right)^2 \kappa_i^+ \frac{1}{1+\left(\frac{\omega}{\kappa_i}\right)^2}  \, , \nonumber
\end{align}
In the limit of fast decay, $\kappa_i\gg \chi_i$,
the dephasing can be treated perturbatively within the Lindblad formalism, Sec.~\ref{sec:LindbladOperators}.
Here one gets the dephasing rate
\begin{eqnarray}
\kappa_{\phi}\approx 8\sum_i \kappa^+_i\left( \frac{\chi_i}{\kappa^-_i}  \right)^2  \, .
\end{eqnarray}
We note that in this limit, 
the contribution from individual two-level systems is then weaker than (maximally) in the strong-coupling case,
$\kappa_{\phi}\ll\sum_i \kappa^+_i$.

\begin{figure}
\includegraphics[width=\columnwidth]{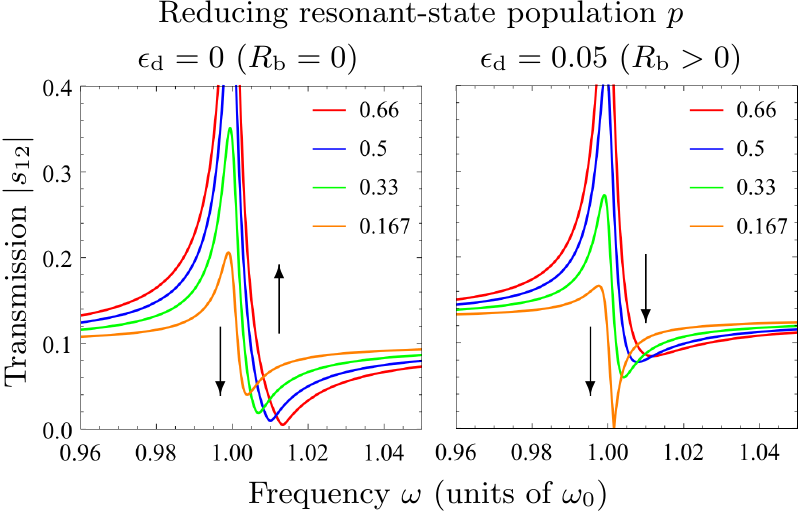}
\caption{Transmission through a cavity with strong dispersive coupling ($\chi\gg\gamma$) to a two-level system that is with probability $p$ in its ground state giving the effective resonance frequency $\omega_0$
(including the dispersive shift).
We assume a weak background transmission $\epsilon=0.05$ and decrease population $p$ as indicated by the plot legends.
When the population is decreased the curve minimum and maximum change as visualized by the arrows.
In the case of dissipation-free background ($\epsilon_{\rm d}=0$)
the reduction in population  decreases the maximum and increases the minimum transmission.
In the case of dissipative background ($\epsilon_{\rm d}=0.05$) the reduction in population  decreases the maximum and the minimum transmission in the shown range of $p$.
The coupling to each transmission line $\gamma= 2\times 10^{-3}\omega_0$.
}\label{fig:ReducedPopulation}
\end{figure}

\subsubsection{Reduced resonant-state population}\label{sec:ReducedPopulation}

In the considered system,
resonance inversion can emerge also without the presence of cavity loss or dephasing, if
the population of the probed state goes essentially below 1. 
A clean example is a regime, where the qubit switching is much slower than the cavity decay to transmission lines, $\kappa^{+/-} \ll \gamma$, so that the effect of hopping is to just modify qubit state populations.
In particular for multi-qubit environments under heating,
it can then occur that the population of the probed state is not 1, but well below it.

Applying Eq.~(\ref{eq:SolutionSeriesCapacitorAmplitude}) and assuming that the transmission through the cavity in the non-resonant state is negligible,  meaning here a background transmission $2(\epsilon-i\epsilon_{\rm d})/(i-2\epsilon+2i\epsilon_{\rm d})$ with probability $1-p$, we get for the average transmission
\begin{align}
s_{12}&= \frac{2( \epsilon-i \epsilon_{\rm d})}{i-2(\epsilon- i \epsilon_{\rm d})}+p\frac{\gamma }{\gamma+i (\omega_0-\omega)} \, .
\end{align}
The effect of reducing probability $p$ is visualized in Fig.~\ref{fig:ReducedPopulation}.
For the dissipation-free background, Fig.~\ref{fig:ReducedPopulation} ($\epsilon_{\rm d}=0$), the interference structure 'straightens',
but for the dissipative background, Fig.~\ref{fig:ReducedPopulation} ($\epsilon_{\rm d}=0.05$), the transmission becomes increasingly skewed and again touches zero.
The curve touches zero with population
\begin{eqnarray}\label{eq:ReducedPopulationMinimum1}
p=1-\frac{\epsilon_{\rm d}}{\epsilon_{\rm d}+2\epsilon_{\rm d}^2+2\epsilon^2} \, 
\end{eqnarray}
and the frequency where this occurs is
\begin{eqnarray}\label{eq:ReducedPopulationMinimum2}
\omega=\omega_0+\frac{\epsilon\gamma}{\epsilon_{\rm d}+2\epsilon_{\rm d}^2+2\epsilon^2} \, .
\end{eqnarray}


\subsubsection{Combined effect}
In a setup with multiple two-level systems, the two above described effects (stochastic artificial-atom swithings and reduced resonant-state population) can coexist.
An approximative model then includes both the finite additional broadening, $\kappa>0$, and the reduced resonant-state population, $p<1$.
Here we have the transmission
\begin{align}\label{eq:Combined}
s_{12}&= \frac{2( \epsilon-i \epsilon_{\rm d})}{i-2(\epsilon- i \epsilon_{\rm d})}+p\frac{\gamma }{\gamma+\kappa/2+i (\omega_0-\omega)} \\
&=\frac{2( \epsilon-i \epsilon_{\rm d})}{i-2(\epsilon- i \epsilon_{\rm d})}+p'\frac{\gamma' }{\gamma'+i (\omega_0-\omega)}  \, .
\end{align}
In the second form, we have defined effective population $p'$ and coupling $\gamma'$,
\begin{align}
p'&=p\frac{\gamma}{\gamma+\kappa/2} \\
\gamma'&=\gamma+\frac{\kappa}{2} \, .
\end{align}
Within these new parameters also Eqs.~(\ref{eq:ReducedPopulationMinimum1}-\ref{eq:ReducedPopulationMinimum2}) are valid.
Furthermore,
the population $p'$ is now the effective spectroscopic signal strength of resonance at frequency $\omega_0$.

\subsubsection{Motional averaging and a limitation of the dephasing rate by dispersive coupling}

\begin{figure}
\includegraphics[width=\columnwidth]{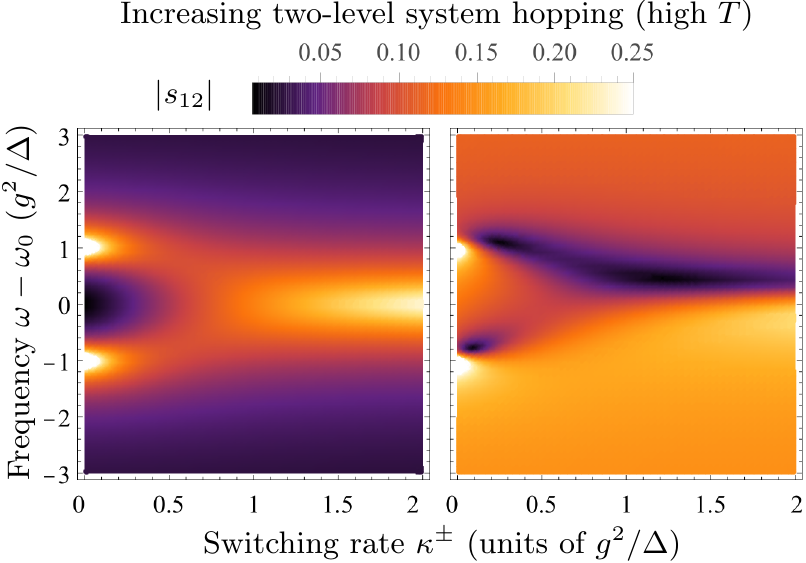}
\caption{Transmission across a linear oscillator with a frequency switching between two values separated by $2g^2/\Delta$.
We consider a situation without ($\epsilon=\epsilon_{\rm d}=0$, left panel) and with ($\epsilon=\epsilon_{\rm d}=0.05$, right panel) a background transmission. The switching rate $\kappa^{\pm}$ is the same for both directions and the dispersive shift is much larger than the cavity linewidth, $g^2/\Delta=13\gamma$, and $\gamma=10^{-3}\omega_0$.
Without a background transmission (left panel) and with weak switching rates, $\kappa^{\pm}<\gamma$, we observe a statistical average of transmission for two equally-populated cavity-frequency states.
When increasing the switching rate, the two transmission peaks merge into a motional-averaged single resonance peak~\cite{Silveri2013}.
In the presence of a weak dissipative background transmission (right panel), the two Fano-type peaks evolve to two dips (when $\kappa^{\pm}\gtrsim\gamma$)
and finally from a single motional-averaged dip to a single tilted peak.
}\label{fig:StarkShiftDensity}
\end{figure}

As an interesting example, showing the diversity of the problem as well as a limitation of the effective cavity dephasing rate,
we consider a cavity frequency that switches with identical hopping rates in the two directions, i.e., $\kappa^+=\kappa^-=\kappa$. Such symmetry appears in the high-temperature limit.

The resulting transmission $\vert s_{12}(\omega)\vert$ as a function of the common hopping rate $\kappa$ and
in the absence of the background is visualized in Fig.~\ref{fig:StarkShiftDensity} (left panel).
Three regimes can be identified. Firstly, if the excitation and decay rates are small compared to the system dynamics, $\kappa\ll\gamma$, the solution for the transmission is a statistical average over two results, corresponding to the two frequencies of the dispersively-shifted cavity. Secondly, when the hopping rate exceeds $\gamma$,
maximum transport reduces significantly due to effective dephasing of cavity resonance frequency.
These are the two regimes considered in the preceding subsections.
Thirdly, when the excitation and decay rates dominate the dispersive shift, $2g^2/\Delta$,
motional averaging emerges~\cite{Silveri2013}, where switching is so fast that only the average value of the cavity frequency is observed.

When a weak parallel transmission with dissipation is then included, Fig.~\ref{fig:StarkShiftDensity} (right panel), the low-hopping rate peaks are changed to skewed Fano-type peaks.
When the hopping rates are increased, the skewed peaks evolve into dips.
A motionally-averaged common dip emerges in the limit of high hopping rates. Finally, a skewed Fano-type peak is recovered in the limit of very high hopping rates.
Our important obervation here  (case $\gamma\ll\chi$) is then that the effective dephasing rate due to stochastic switching
of dispersively-coupled two-level systems
is limited by the dispersive coupling, i.e.,  $\kappa_{\phi}< \chi$.

We note that a similar line-shape transformation can also occur without the presence of dispersive-shift hopping, but,
through other phenomena reducing the cavity transmission, such as photon blockade when increasing drive power~\cite{Bishop2009,Bishop2010,Ping2018}. 
Such higher drive amplitudes can be useful for quantum sensors~\cite{Schneider2018}
or quantum simulation~\cite{Braumuller2017,Leppakangas2018}.




\subsection{Higher levels of artificial atoms}\label{sec:HigherEnergyLevels}

Higher excited states of artificial atoms can influence the observed peak transformation.
A Hamiltonian that can be used to account for higher excited states is given in Sec.~\ref{sec:DispersiveManyLevels}.

For transmon artificial atoms, a simple estimate for the population of excited states at a given temperature can be  made.
Here, the anharmonicity $E_C$ is usually 5-10 percent of the lowest energy-level spitting $\hbar\Omega$.
This means that for moderate temperatures the populations can be assumed to be the ones of a harmonic oscillator,
\begin{align}
P_n &\approx \frac{1}{1+N}\left(\frac{N}{N+1}\right)^2 \\
N &= \frac{1}{\exp\beta\Omega-1} \, .
\end{align}
This formula can then be used to estimate how significantly the higher energy levels contribute to cavity dephasing.
In our experiment, the local temperature varies in a range that implies $P_1\lesssim 0.2$, giving
$P_2\lesssim 0.05$. This means that higher excited states are not significantly populated and
their contribution to the analyzed effects stays negligible.
For simplicity, we then neglect their contribution from the theoretical model. It should however be noted that for
transmons, the second excited energy-levels strongly modify the dispersive shifts of the first excited states, see Sec.~\ref{sec:DispersiveManyLevels}.



\begin{figure}
\includegraphics[width=\columnwidth]{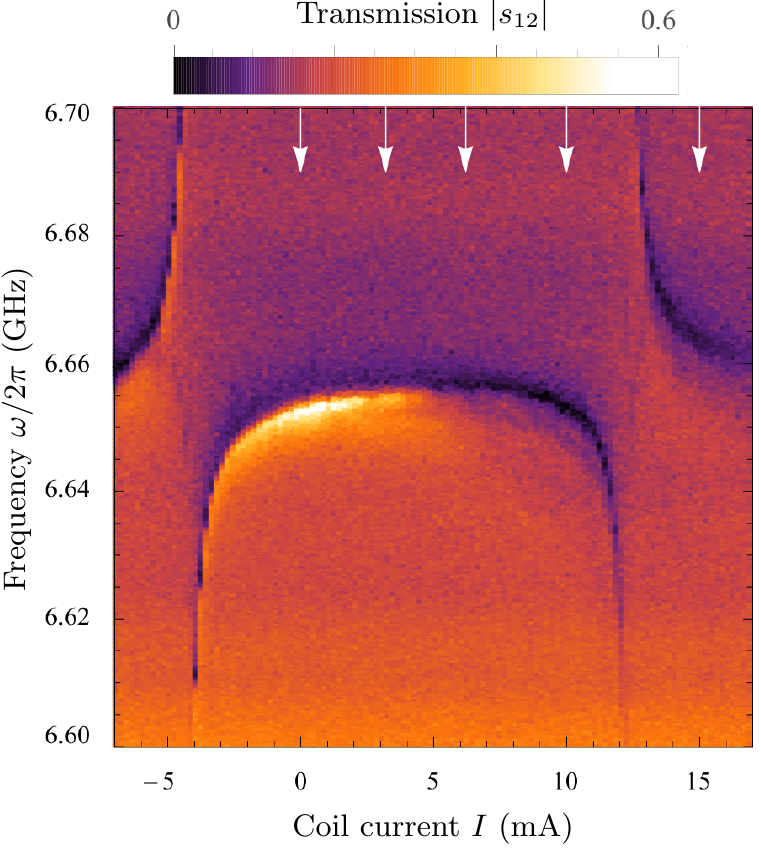}
\caption{
Measured microwave transmission $\vert s_{12} \vert$ through a coplanar cavity coupled to eight transmons as a function of coil current~$I$, used to tune one transmon, and drive frequency~$\omega$.
The resonance position is periodic as a function of the current (transmon inductance), whereas the line shape transforms from a peak to a dip when current $\vert I \vert$ is increased.
Transmission curves at pointed current bias points are shown in Fig.~\ref{fig:Comparison11}.
}\label{fig:Comparison1}
\end{figure}

\section{Experiment}\label{sec:Experiment}

Our experiment includes 8~transmon artificial atoms embedded in a driven coplanar microwave resonator~\cite{Ping2018}.
The equivalent circuit model is shown in Fig.~\ref{fig:setup2}(a) with the cavity-atom coupling scheme of
Fig.~\ref{fig:setup2}(c).
The energy levels of the transmons are individually tunable by local magnetic fluxes applied
across Josephson inductors realized in a SQUID geometry.
The magnetic fluxes are created by DC bias currents through 8~nearby coils.
The energy levels of the multi-atom environment affect the effective cavity resonance frequency, which
is probed by measuring the microwave transmission $s_{12}$ through the cavity.
Further technical details of the experiment are given in Appendix~E and in Ref.~\cite{Ping2018}.

\subsection{Uncalibrated system: Resonance inversion by heating}\label{sec:Uncalibrated}
An example of measured microwave transmission $s_{12}$ is shown in Fig.~\ref{fig:Comparison1}.
Here we probe the system in the neighborhood of the dispersively-shifted cavity $n=1$ mode at base temperature $T\approx 20$~mK.
This mode has pure frequency $\omega_0/2\pi=6.674$~GHz.
We are in the weak-drive limit and the data is normalized according to maximal transmission at high powers.
We sweep single coil current $I$ to tune one transmon. All other coil currents are zero.


In Fig.~\ref{fig:Comparison1},
we observe an avoided energy-level crossing between the cavity and the tuned transmon.
The observed resonance-frequency variation is periodic as a function of current $I$, as expected from SQUID flux periodicity.
An additional random offset flux threads the SQUID loops leading to an offset from the symmetry-point $I=0$.
An unexpected feature is the reduction of on-resonance
transmission with increasing $\vert I \vert$, with a line shape changing from a peak to a dip.
This is a general feature of the experiment in different biasing conditions and cool downs.

\begin{figure} 
\includegraphics[width=\columnwidth]{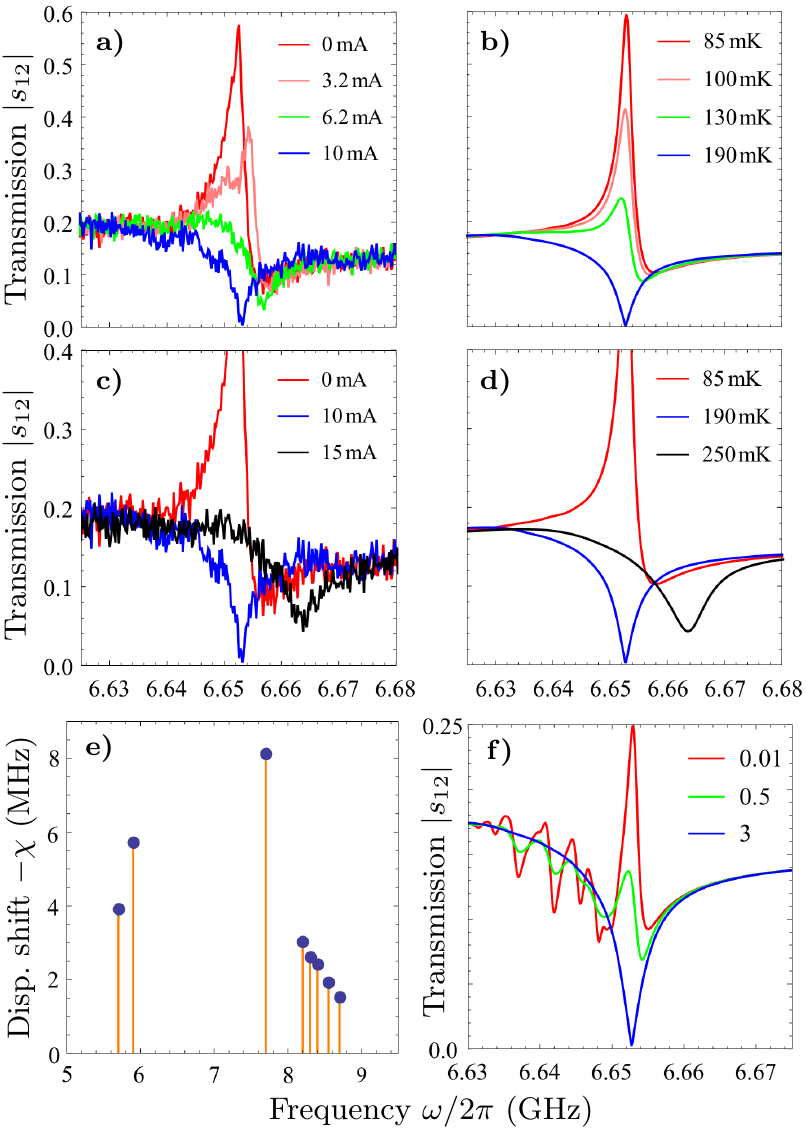}
\caption{
a) Experimental data of Fig.~\ref{fig:Comparison1} for indicated values of coil current $I$, showing a reduction of maximum transmission and transition from a peak to a dip when current is increased.
b) The change of the line shape as given by master-equation (ME) simulation of cavity coupled dispersively to 8~two-level systems.
When temperature is increased the transmission minimum and maximum reduce.
c-d) A comparison for wider range of applied currents and temperatures.
e) The transmon two-level system energies and equivalent dispersive shifts used in the ME simulation.
f) A ME simulation with $T=190$~mK and varying common qubit decay $\kappa_i$.
The curve keys give equivalent decay rates at $T=0$ in units of the resonator coupling $\gamma=2\pi\times 0.7$~MHz. The curve minimum decreases with increasing $\kappa_i$.
Fitting such curve to experimental data (at known temperatures) can be used to estimate the average decay rate of qubits.
}\label{fig:Comparison11}
\end{figure}

The experimental data at several current biases is compared to master-equation simulations in Fig.~\ref{fig:Comparison11}(a-d).
At these bias points the cavity-transmon coupling is dispersive.
We simulate the system as a linear resonator coupled to eight two-level systems.
An important feature of the experiment is the large background transmission, here $\vert s_{12} \vert \lesssim 0.2$.
This background transmission corresponds to model parameter $\epsilon-{\rm i}\epsilon_{\rm d}=0.062-{\rm i}0.06$ and
was measured separately when the cavity resonance frequency was tuned far away (by tuning qubits) and by using the connection~(\ref{eq:EpsilonInverse}).
The background transmission originates most probably in a crosstalk between the transmission lines, flux-bias lines, and the sample-box bonding. For more details see Appendix~E.

In master-equation simulations,
we use the hypothesis that DC current causes heating of 
the cavity-atom system.
As a result, stochastic excitations (and decays) of the heated two-level systems make the resonator frequency
fluctuate. The assumption of elevated temperature is supported by
additional experiments on the base-temperature dependence, which imply that for large coil currents $T\sim 0.2$~K.
Furthermore, a spectroscopy of higher exciation manifolds described in Ref.~\cite{Ping2018} implied $T\sim 0.15$~K.
At such temperatures, thermal populations of higher excited-states of transmons are however small:
it is consistent to decribe transmons as two-level systems.
The base temperature itself was experimentally found to depend on the coil currents only weakly.
The origin of the observed local heating is most probably resistive heating in the Copper-PCB leads in combination with
a weak thermalization of the chip to the sample box.
We also note that a qualitatively similar dip in microwave transmission can be caused by a direct reflection from two-level systems~\cite{Hoi2012}.
However, in this case the effect of a temperature would be opposite, i.e., zero temperature would give the (deepest) dip.
This process can also be ruled out by tuning the qubits, since such mechanism would work only at qubit frequencies,
whereas the observation is that the dip appears at the cavity frequency.

In master-equation simulations, we use measured cavity-transmon couplings $g_i/2\pi\approx 110$~MHz and anharmonicities $E_C\approx 410$~MHz~\cite{Ping2018}.
The coupling of the cavity to transmission lines as given by high-power transmission is
$\gamma=2\pi\times 0.7$~MHz, in accordance with the value obtained from microwave circuit simulations. 
The used qubit energies are distributed around the resonance frequency, with a restriction that total
dispersive shift $20$~MHz is reproduced.
Therefore here only an estimation of the qubit frequencies can be made, following from that
the experimental control was not calibrated (a calibrated situation is studied below).
In our system, only
this uncalibrated situation however allows for observing the heating induced inversion of the line shape,
since after the calibration the system is always at an elevated temperature.
The distribution of frequencies (and thereby dispersive shifts) used in the simulation is shown in Fig.~\ref{fig:Comparison11}(e).
It should also be noted that in reality
the qubit frequencies also shift slightly when magnetic flux across the target qubit is being changed,
which is not accounted for by the simulation. An  exception is the last transmission curve ($T=250$~mK),
where the resonance curve is shifted 10~MHz upwards to account for the changed resonator frequency-shift
due to tuning of the target qubit.

In Fig.~\ref{fig:Comparison11}(f), we visualize how the peak form can also be used to estimate the $T_1$ times of the qubits.
Here, we fix the system temperature to $T=190$~mK. The higher the decay rate, the higher the dephasing rate becomes.
Simultaneously, the fine structure smoothens out.
The observed peak depths and forms, together with assuming temperatures obtained from other observations,
lets us to conclude that an average qubit decay rate is between $2-3\gamma$, which means $T_1\sim 50-80$~ns.
It should be noted that the $T_1$ time measured for a single-qubit sample was essentially longer, $T_1\sim 500$~ns.
Such direct measurement of $T_1$ in the 8~qubit sample was however not possible.
The relatively short theoretical value is further supported by simulations when qubits are tuned on resonance with the cavity, see Sec.~\ref{sec:Calibration}.

\begin{figure} 
\includegraphics[width=\columnwidth]{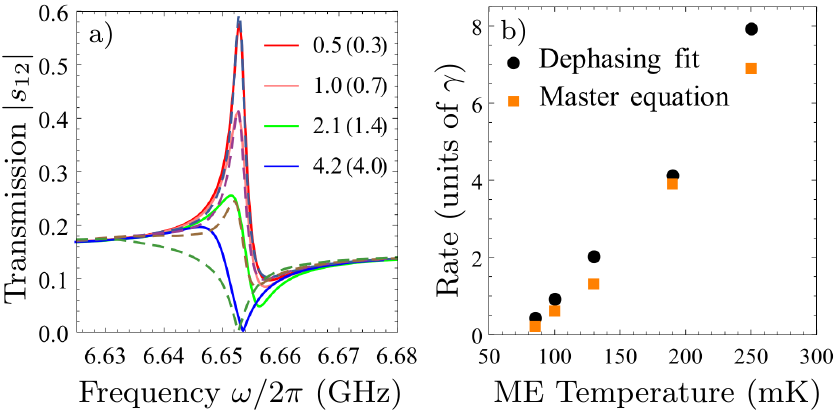}
\caption{
a) Transmission across a linear cavity with internal dephasing $\kappa_{\phi}$ (solid lines) compared to the master equation (ME) simulation of 8 heated two-level systems (dashed lines).
In the linear cavity model, the effect of heated multi-qubit environment is accounted for by the introduced phenomenological dephasing rate $\kappa_{\phi}$, given in the units of cavity coupling $\gamma=2\pi\times 0.7$~MHz. The parenthesis value corresponds to the sum of excitation rates of all qubits in the ME model, $\kappa_{\Sigma}=\sum_{i=1}^8\kappa_i^+$.
The curve minima decrease with increasing $\kappa$.
b) A comparison between the phenomenological dephasing rate $\kappa_{\phi}$
and the corresponding master-equation rate $\kappa_\Sigma$. Here the temperature (x-axis)
is defined by the master-equation simulation. 
}\label{fig:Comparison111}
\end{figure}

In Fig.~\ref{fig:Comparison111}(a), we additionally consider fitting the peak inversion by
the model of a linear resonator with increasing internal dephasing rate  $\kappa_{\phi}$ (or equivalently internal loss rate  $\kappa/2$).
Such a simplified model accounts for transitions in the heated two-level system ensemble as an effective cavity dephasing rate $\kappa_{\phi}$.
We obtain that an increase of dephasing rate gives a qualitatively similar
line-shape transformation as the full master-equation model with heating (the latter plotted here as dashed lines).
Furthermore, the dephasing rates are close to the sums of the excitation rates of two-level systems, see Fig.~\ref{fig:Comparison111}(b) [see also Eq.~(\ref{eq:DephasingConnection2})].


\subsection{Calibrated system: Transmission in different backgrounds}\label{sec:Calibration}

\begin{figure} 
\includegraphics[width=\columnwidth]{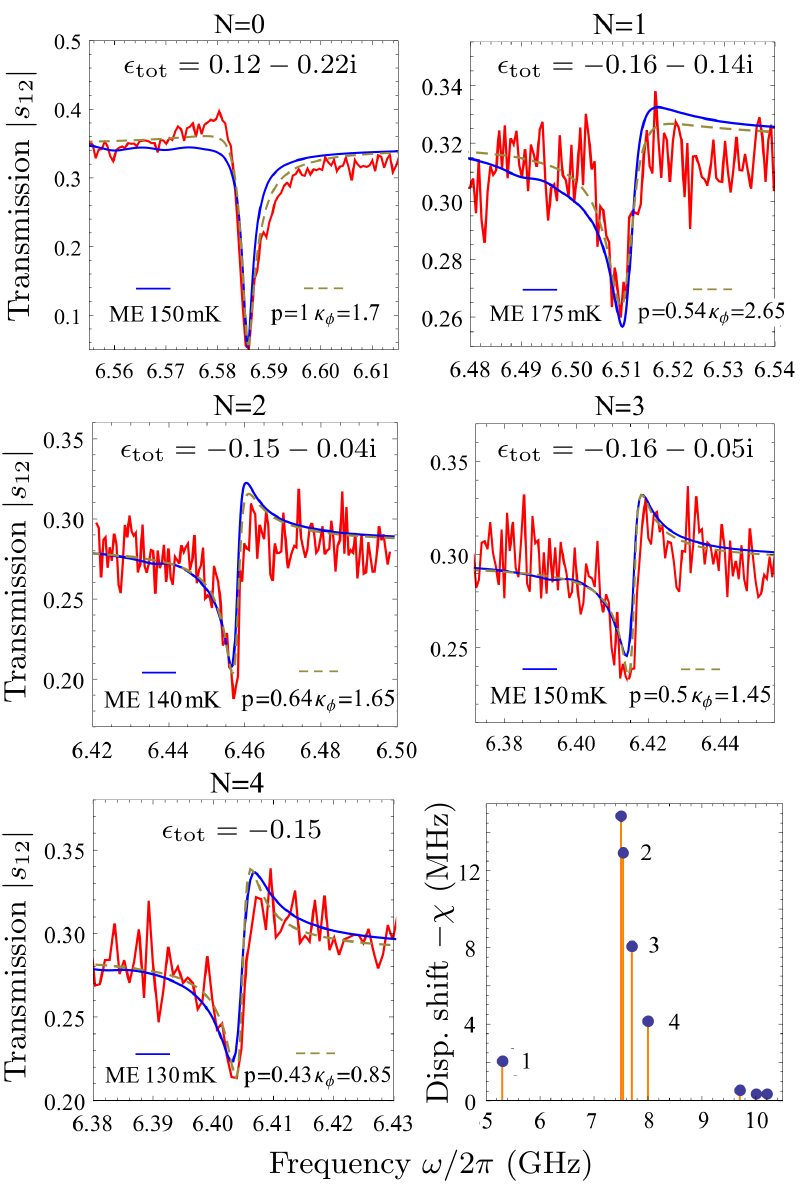}
\caption{ Measured
transmission when different number of qubits (from $N=0$ to $N=4$) are on resonance with the cavity.
The background parameter, $\epsilon_{\rm tot}=\epsilon-{\rm i}\epsilon_{\rm d}$, is different for each $N$.
The master-equation (ME) simulation treats temperature $T$ as a free fitting parameter.
A fit based on the model of a linear resonator with resonant-state population $p$ and
dephasing rate $\kappa_\phi$, Eq.~(\ref{eq:Combined}), is plotted as dashed lines.
(Bottom right) Two-level system frequencies and dispersive shifts before the on-resonance tuning, i.e., case $N=0$.
The two-level systems  tuned on resonance are marked by numbers $1\rightarrow 4$.
}\label{fig:Comparison1111}
\end{figure}

In this section, 
we study transmission while tuning the effective resonance frequency of the cavity.
Since the background transmission is in our system frequency dependent, see Appendix~E,
the change of the cavity frequency corresponds to changing the background.

We consider here the case of a fully calibrated system,
allowing for the control of all eight transmons and thereby the dressed-cavity frequency.
In comparison to the uncalibrated system studied in Sec.~\ref{sec:Uncalibrated},
here a significant total current flows always in the flux-bias lines,
leading to an elevated temperature of the system.
We then cannot reproduce the resonance inversion effect in this operation scheme (transformation from a peak to a dip by heating).

In Fig.~\ref{fig:Comparison1111}, we consider transmission
when different number of qubits (from $N=0$ to $N=4$) are on resonance with the cavity mode.
We study transmission around the lower resonance peak of the splitted cavity-qubit peak~\cite{Ping2018}.
The equivalent background parameters, $\epsilon_{\rm tot}=\epsilon-{\rm i}\epsilon_{\rm d}$,
are also shown in Fig.~\ref{fig:Comparison1111} (from $N=0$ to $N=4$) and were
determined from off-resonance data.
The experimental data are compared to
master-equation simulations with calibration-determined qubit frequencies and dispersive shifts, shown
in Fig.~\ref{fig:Comparison1111} (bottom right).
We have fixed the qubit decays to $\kappa_i=3\gamma$, in accordance with the dispersive-regime analysis, Sec.~\ref{sec:Uncalibrated}. We use temperature as a free fitting parameter. We
find a good agreement between the theory and experiment by varying simulation
temperature between 130-175~mK. This range of temperatures is consistent with the analysis of Sec.~\ref{sec:Uncalibrated}.
We have then obtained a reliable estimate for the local temperature at each bias point.


In Fig.~\ref{fig:Comparison1111}, we additionally
fit the observed line shapes with the model of a linear resonator with (resonant-state) population $p$ and
phenomenological dephasing rate $\kappa_\phi$, see Sec.~\ref{sec:JaynesCummingsOscillatorDecoherence}.
The fit is similar to the one obtained from the master-equation simulation.
We find that the population $p$ reduces when the first qubit is brought on resonance.
This is since on-resonance
transition frequencies between nearby photon numbers are not anymore approximately constant:
heating of the resonator takes the system away from the subspace providing the transmission resonance.
On the other hand, bringing dispersively-coupled qubits on resonance decreases broadening,
since their stochastic-hopping induced dephasing is removed. 
This leads to tendency of reducing  $\kappa_\phi$ when $N$ increases. It also leads to reduction of $p$
and finally to disappearance of the resonance behind the noise when $N>5$~\cite{Ping2018}.

We also note that
the lineshapes are generally affected by pure dephasing of artificial atoms. In the considered setup,
however, the width given by average qubit lifetimes already reproduces the observed linewidths well.
This means that here dephasing of qubits is dominated by the fast qubit decay, $T_1\sim 50-80$~ns.
Such line-width fitting, together with the dispersive-regime line-shape analysis, demonstrates how $T_1$ and $T_2$ times can,
in principle, be estimated independently, without resorting to time-domain measurements.

\section{Conclusions and discussion}\label{sec:Conclusions}

In this work
we have investigated Fano resonances in microwave transmission across a two-sided cavity coupled to multiple artificial atoms and
in the presence of a microwave background. The background was effectively formed by the artificial-atom control circuitry.
Present and future microwave  quantum-information applications wrap together high number of qubits and control lines
in a finite sized chip and a sample box, where Fano resonances can easily occur.
We have then studied in detail how to account for such resonances in most common quantum-microwave models
and particularly how the line shapes of the energy levels can connect to
dissipation and fluctuations in such systems,
helping to better understand possibly complex spectroscopic data of this type of devices.

An important theoretical result was that the background does not necessarily affect the equation of motion of the cavity and artificial atoms. Instead, it can be included to theoretical results obtained, for example, from the well-known Jaynes- and Tavis-Cummings models afterwards by applying modified linear boundary conditions. Its effect can also be substracted out (and undone) straightforwardly from measurement data~\cite{Ping2018}.
This property remains to be valid also for time-dependent fields (measurement pulses), higher drive powers, and beyond the two-level system and dispersive-regime approximations.
Furthermore,
we showed how temperature of the multi-qubit environment can be estimated from the line shape of the resonator,
as well as how average $T_1$ and $T_2$ times of qubits can be determined without doing direct time-domain measurements
(or independently of them).

It should also be noted that all dephasing mechanisms of superconducting microwave resonators due to coupling to spurious two-level systems~\cite{Gao2007,Vissers2012,Burnett2014,deGraaf2018} are not yet fully understood.
The results obtained here for the behavior of Fano-type resonances
also apply to studies of such systems generally. Systems as described here can also be used as quantum simulators
to experimentally study the involved physical phenomena.

\acknowledgements
This work was supported by the European Research Council (ERC) under the Grant Agreement 648011, Deutsche Forschungsgemeinschaft (DFG) within Project No.~WE4359/7-1, the Initiative and Networking Fund of the Helmholtz Association, the China Scholarship Council (CSC), and Studienstiftung des deutschen Volkes.
We also acknowledge support provided by the Initiative and Networking Fund of the Helmholtz Association, within the Helmholtz Future Project Scalable solid state quantum computing. This work was also partially supported by the Ministry of Education and Science of the Russian Federation in the framework of the Program to Increase Competitiveness of the NUST MISIS, contract no. K2-2017-081.


\appendix

\section{Microwave scattering in a linear circuit}
In this appendix, we apply a classical circuit model to study microwave transmission and reflection in our system.
This model assumes that the (whole) system can be described as a set of lumped circuit elements.
The model accounts for superconducting artificial atoms as $LC$ resonators.

In this approach,
we derive scattering properties by applying Kirchhoff rules at the cavity boundaries.
For this, we first identify
the total voltage and total current due to forward (in) and backward (out) propagating fields at the two sides of the cavity.
On the left-hand side these are~\cite{Pozar}
\begin{eqnarray}
V^{\rm L} &=& V^{\rm L}_{\rm in}+V^{\rm L}_{\rm out}\label{eq:BoundaryLeft1} \\
I^{\rm L} &=& \frac{V^{\rm L}_{\rm in}}{Z_0}-\frac{V^{\rm L}_{\rm out}}{Z_0} \label{eq:BoundaryLeft2} 
\end{eqnarray}
and on the right-hand side
\begin{eqnarray}
V^{\rm R} &=& V^{\rm R}_{\rm in}+V^{\rm R}_{\rm out}\label{eq:BoundaryRight1} \\
I^{\rm R} &=& -\frac{V^{\rm R}_{\rm in}}{Z_0}+\frac{V^{\rm R}_{\rm out}}{Z_0} \label{eq:BoundaryRight2} \,.
\end{eqnarray}
These variables are Fourier components of the total propagating field, e.g., $V^{\rm L}=V^{\rm L}(\omega)$. 
The reflection $s_{11}$ and the transmission $s_{12}$ amplitudes  are here
\begin{eqnarray}
s_{11}^*&=&  \frac{V_{\rm out}^{\rm L}}{V_{\rm in}^{\rm L}}  \\
s_{12}^*&=&  \frac{V_{\rm out}^{\rm R}}{V_{\rm in}^{\rm L}} \, .
\end{eqnarray}
We assume here $V_{\rm in}^{\rm R}=0$.
The complex conjugation is needed here in comparison to Eqs.~(\ref{eq:ReflectionOperatorDefinition1}-\ref{eq:ReflectionOperatorDefinition2}) since the impedance treatment assumes implicitly a time dependence $\sim V(\omega) e^{i\omega t}$, which is opposite to the time dependence of annihilation operators $\sim\hat a(\omega)e^{-i\omega t}$.

We first consider the case of parallel impedance  ($C_{\rm c}=0$).
This gives two boundary conditions, which state current conservation and voltage drop across the impedance $Z_{\rm b}(\omega)$,
\begin{align}
\frac{V^{\rm L}_{\rm in}}{Z_0}-\frac{V^{\rm L}_{\rm out}}{Z_0} &=  -\frac{V^{\rm R}_{\rm in}}{Z_0}+\frac{V^{\rm R}_{\rm out}}{Z_0} \\
Z_{\rm b}\left( \frac{V^{\rm L}_{\rm in}}{Z_0}-\frac{V^{\rm L}_{\rm out}}{Z_0}  \right)   &= V_{\rm in}^{\rm L} + V_{\rm out}^{\rm L}-\left( V_{\rm in}^{\rm R} + V_{\rm out}^{\rm R}\right) \, .
\end{align}
Using $V_{\rm in}^{\rm R}=0$ the solution is
\begin{eqnarray}
s_{11}^* &=&  \frac{1}{1+\frac{2Z_0}{Z_{\rm b}} } \label{eq:SolutionParallel1} \\
s_{12}^* &=&  \frac{\frac{2Z_0}{Z_{\rm b}}}{1+\frac{2Z_0}{Z_{\rm b}}}\label{eq:SolutionParallel2}  \, .
\end{eqnarray}

We can now study more detailed the effect of dissipation in the parallel channel. In the main text this was done by introducing the imaginary part, $\epsilon_{\rm tot}=\epsilon-{\rm i}\epsilon_{\rm d}$.
In the impedance approach the equivalent parameter is $i\epsilon_{\rm tot}=Z_0/Z^*_{\rm b}(\omega)=Z_0Z_{\rm b}/\vert Z^*_{\rm b}(\omega)\vert^2$.
Dissipation is included, for example, by changing the impedance
from $Z_{\rm b}(\omega)=i\omega L_{\rm b}\omega$ to  $Z_{\rm b}(\omega)=i\omega L_{\rm b}\omega+R_{\rm b}$, with a series resistance $R_{\rm b}>0$.
This approach also shows that a dissipative part renormalizes the reactive term $\epsilon$. This is studied further below.
We note that in a direct comparison of Eqs.~(\ref{eq:SolutionParallel1}-\ref{eq:SolutionParallel2}) to $s_{11}$ and $s_{12}$  given in the main text, Eqs.~(\ref{eq:SolutionSeriesCapacitorAmplitude}-\ref{eq:SolutionSeriesCapacitorAmplitudeReflection}), an overall minus-sign difference appears due to different definition of scattering-state phases, for more details see Appendix~B.

Similarly, we can construct boundary conditions for arbitrary cavity couplings $Z_{\rm c1}=1/i\omega C_{\rm c1}$ and $Z_{\rm c2}=1/i\omega C_{\rm c2}$.
Here we allow for different coupling capacitance of the cavity to the left-hand side ($C_{\rm c1}$) and the right-hand  side ($C_{\rm c2}$) TLs.
We then consider Kirchhoff equations for the input and output fields as well for the voltage on the island between capacitances $C_{\rm c1/2}$, which we mark now $V$.
The resulting equations have the form
\begin{widetext}
\begin{eqnarray}
\left(
\begin{matrix}
\frac{1}{Z_0} + \frac{1}{Z_{\rm c1}(\omega)} + \frac{1}{Z_{\rm b}(\omega)} & -\frac{1}{Z_{\rm c1}(\omega)} & -\frac{1}{Z_{\rm b}(\omega)}\\
-\frac{1}{Z_{\rm c1}(\omega)} & \frac{1}{Z(\omega)} + \frac{1}{Z_{\rm c1}(\omega)} + \frac{1}{Z_{\rm c2}(\omega)} & -\frac{1}{Z_{\rm c2}(\omega)}       \\
\frac{1}{Z_{\rm b}(\omega)} & \frac{1}{Z_{\rm c2}(\omega)} & -\frac{1}{Z_0} - \frac{1}{Z_{\rm c2}(\omega)} - \frac{1}{Z_{\rm b}(\omega)}
\end{matrix}
\right)
\left(
\begin{matrix}
V_{\rm out}^{\rm L} \\
V \\
V_{\rm out}^{\rm R} 
\end{matrix}
\right)=
\left(
\begin{matrix}
\frac{1}{Z_0}-\frac{1}{Z_{\rm c1}(\omega)}-\frac{1}{Z_{\rm b}(\omega)} \\
\frac{1}{Z_{\rm c1}(\omega)} \\
-\frac{1}{Z_{\rm b}(\omega)} 
\end{matrix}
\right)V_{\rm in}^{\rm L} \, ,\nonumber
\end{eqnarray}
\end{widetext}
The answer for the output fields and for the island voltage as a function of the input $V_{\rm in}^{\rm L}$ can then be found easily by a matrix inversion.
For the considered symmetric coupling, $C_{{\rm c}1}=C_{{\rm c}1}=C_{{\rm c}}$, the analytical solution is
\begin{align}
s_{11}^* &=  \frac{ 2Z_0  \left[Z_{\rm c}^2 +Z(2Z_{\rm c}+Z_{\rm b})   \right]  }{(2Z+Z_0+Z_{\rm c})\left[Z_0(2Z_{\rm c}+Z_{\rm b})+Z_{\rm c}Z_{\rm b}\right]}    \\
s_{12}^* &= \frac{  Z_{\rm c}Z_{\rm b}( 2Z+Z_{\rm c}) -Z_0^2(2Z_{\rm c}+Z_{\rm b}) }{(2Z+Z_0+Z_{\rm c})\left[Z_0(2Z_{\rm c}+Z_{\rm b})+Z_{\rm c}Z_{\rm b}\right]}  \, .
\end{align}

\begin{figure}
\includegraphics[width=\columnwidth]{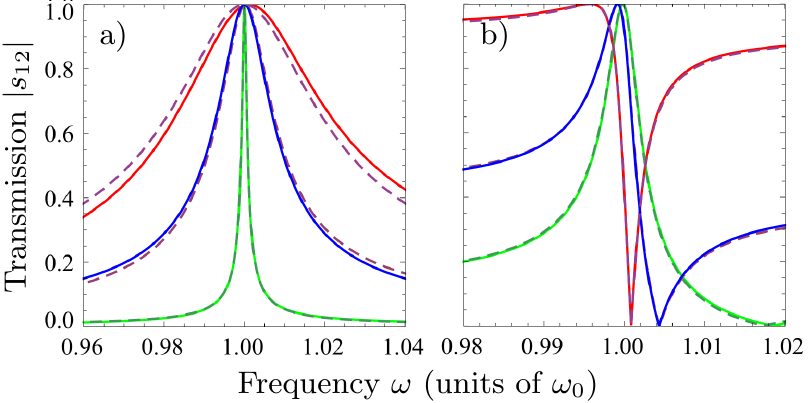}
\caption{Microwave transmission amplitude $\vert s_{12}\vert$ across a linear two-sided cavity.
We do comparison between the single-mode treatment (solid lines) and impedance approach (dashed lines).
The cavity is described by an $LC$ oscillator.   
(a) Transmission peak without the presence of background transmission with $C_{\rm c}/C=$0.05, 0.2, 0.4. 
In the single-mode approximation this corresponds to $\gamma=(0.5, 8, 32)\times 10^{-3}\omega_0$. The resonance peak widens with increasing $C_{\rm c}$ or $\gamma$.
The single-mode treatment is here a good approximation and becomes exact in the limit $C_{\rm c}/C\rightarrow 0$.
(b) Transmission for $C_{\rm c}/C=0.1$ with increasing dissipationless background transmission corresponding to $\epsilon=0.05$, 0.2, 1.0. Here, the single-mode treatment is a good approximation for all strengths of the background transmission.
The off-resonance transmission increases when increasing $\epsilon$.
}
\end{figure}\label{fig:Theory_Comparison_Linear_Heisenberg2}

A comparison between the single-mode treatment of the main text and the impedance approach considered here is shown in Fig.~11.  
We consider dissipationless background transmission.
Here, the impedance model is exact whereas the single-cavity-mode model used in the main text is an approximation.
We find that if $C_{\rm c}\ll C$, the single-cavity-mode model works well for all strengths of parallel transmission, i.e., for all values of $\epsilon$.

A comparison in the case of dissipative background is shown in Fig.~12. 
We determine $\epsilon_{\rm tot}=\epsilon-i\epsilon_{\rm d}$  from identification
$i\epsilon_{\rm tot}=Z_0/Z^*_{\rm b}(\omega)=Z_0Z_{\rm b}/\vert Z^*_{\rm b}(\omega)\vert^2$ by using
$Z_{\rm b}(\omega)=i\omega L_{\rm b}\omega+R_{\rm b}$ and insert this parameter into the single-mode model.
We find again that for $C_{\rm c}\ll C$
the exact linear solution and the approximative single-cavity-mode model are practically the same.

\begin{figure}
\includegraphics[width=\columnwidth]{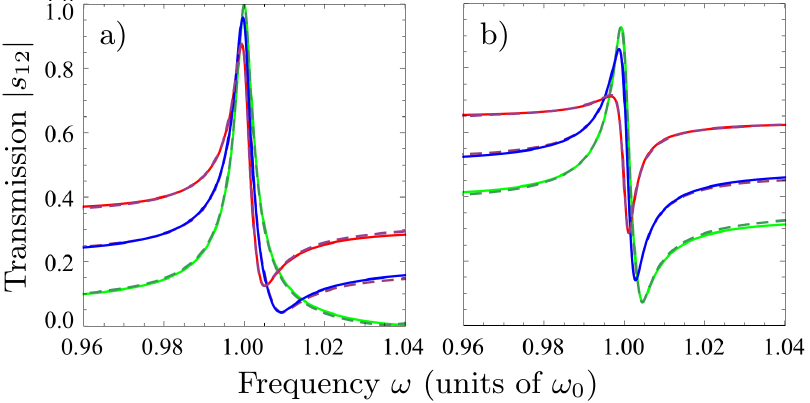}
\caption{Microwave transmission amplitude $\vert s_{12}\vert$ across a linear two-sided cavity
for varying background inductance $L_{\rm b}$ and fixed background resistivity $R_{\rm b}$.
We do comparison between the single-mode treatment (solid lines) and impedance approach (dashed lines) of the cavity.
(a) Transmission for $C_{\rm c}/C=0.1$ with reducing inductance with fixed resistivity.
The equivalent $\epsilon$-parameters are $\epsilon=0.027,0.10,0.18$ and dissipative part $\epsilon_{\rm d}=0.0015, 0.023, 0.08$.
The off-resonance transmission increases when increasing $\epsilon$ (reducing inductance).
(b) Transmission for $C_{\rm c}/C=0.1$ with reducing inductance further with fixed resistivity. The equivalent $\epsilon$-parameters are $\epsilon=0.21, 0.32, 0.50$ and dissipative part $\epsilon_{\rm d}=0.046, 0.12, 0.55$.}
\end{figure}\label{fig:Theory_Comparison_Linear_Heisenberg_Dissipation}

\section{Deriving boundary conditions and Heisenberg equations of motion}

The boundary conditions and Heisenberg equations of motion can be derived by starting from a Lagrangian approach for
an open transmission line interacting with a cavity. Here, we discretize the transmission line
to elements of length $\delta x$ with capacitance to ground $\delta x C'$ and inductance $\delta x L'$ in between.
We apply this approach to the case of free cavity (no transmons), whereas generalization to
the case of cavity embedding multiple transmons is straightforward.


In this approach, the total Hamiltonian of the system can be derived to be
\begin{eqnarray}
H_{\rm tot}&=& H_{\rm R}+H_{\rm L}+H_{\rm res}+H_{\rm int} \, .
\end{eqnarray}
The Hamiltonian that describes the right-hand open transmission line is
\begin{eqnarray}
H_{\rm R}&=&\sum_{r=2}^{\infty} \frac{ Q_r^2}{2\delta xC'} +\sum_{r\geq 2}^\infty\frac{(\Phi_r-\Phi_{r-1})^2}{2L'\delta x} \\
&+&  \frac{QQ_1}{C} +\frac{Q_1^2}{2C_{\rm s}} \, . \nonumber
\end{eqnarray}
Indices $r\in [1,\infty]$ refer to nodes of the discretized transmission line, value $r=1$ corresponds to the node next to the cavity. 
Variables $\Phi_r$ and $Q_r$ correspond to a magnetic flux and charge at node $r$ and an effective series capacitance is defined as $1/C_{\rm s}=1/C_{\rm c}+1/C$.  The flux (charge) variable of the resonator is $\Phi$ ($Q$). 
Similarly for the left-hand side Hamiltonian,
\begin{eqnarray}
H_{\rm L} &=&\sum_{l=-\infty}^{-2} \frac{Q_l^2}{2\delta xC'} +\sum_{l\leq -1}\frac{(\Phi_l-\Phi_{l-1})^2}{2L'\delta x}  \\
&+&  \frac{QQ_{-1}}{C} +\frac{Q_{-1}^2}{2C_{\rm s}} \, . \nonumber
\end{eqnarray}
A Hamiltonian that describes the in-line resonator is
\begin{eqnarray}
H_{\rm res}= \frac{Q^2}{2C}+ \frac{ \Phi^2}{2L}\, .
\end{eqnarray}
The inductance $L$ and capacitance $C$ are defined in Fig.~\ref{fig:setup}(b).
Finally, the direct interaction between the two transmission lines (background) is described by
\begin{eqnarray}
H_{\rm int}= \frac{\left(\hat \Phi_{1}-\hat \Phi_{-1}\right)^2}{2L_{\rm b}}+  \frac{Q_{-1}Q_1}{C}  \, ,
\end{eqnarray}
where $L_{\rm b}$ is the assumed inductive coupling through the background. Also a direct-coupling-type term through the resonator appears.

The Heisenberg equations of motion for the transmission lines in the limit $\delta x\rightarrow 0$ result in a wave equation whose solution can be written as in Eq.~(\ref{eq:MW1}).

The Heisenberg equations at the node $r=1$ are
\begin{eqnarray}
\hat{\dot{ \Phi}}_{1}(t) &=&  \frac{\hat Q}{C}+\frac{\hat Q_{-1}}{C}+ \frac{\hat Q_{1}}{C_{\rm s}} \label{boundary1} \\
\hat{\dot{ Q}}_{1}(t) &=&\frac{1}{L'}\frac{\partial \hat \Phi(x=0^+,t)}{\partial x} + \frac{\hat \Phi_{-1}-\hat \Phi_{1}}{L_{\rm b}} \,. \label{boundary2}
\end{eqnarray}
Similarly for the left-hand side ($l=-1$)
\begin{eqnarray}
\hat{\dot{ \Phi}}_{-1}(t) &=&  \frac{\hat Q}{C}+\frac{\hat Q_{1}}{C}+ \frac{\hat Q_{-1}}{C_{\rm s}} \label{boundary11} \\
\hat{\dot{ Q}}_{-1}(t) &=&-\frac{1}{L'}\frac{\partial \hat \Phi(x=0^-,t)}{\partial x} + \frac{\hat \Phi_{1}-\hat \Phi_{-1}}{L_{\rm b}} \,. \label{boundary22}
\end{eqnarray}
Eq.~(\ref{boundary2}) is satisfied by
\begin{eqnarray}\label{eq:SolutionQ1}
\hat Q_1(t)&=&\sqrt{\frac{\hbar }{2\omega_{0}Z_{0}}}\left[-\hat a^{\rm R}_{\rm  in}(t)+ a^{\rm R}_{\rm  out}(t)\right] \\
&+& \frac{i}{\omega_0L_p}\sqrt{\frac{\hbar Z_0 }{2\omega_0}}\left[\hat  a^{\rm L}_{\rm  in}(t)+ a^{\rm L}_{\rm  out}(t)\right]\nonumber\\
&-&\frac{i}{\omega_0L_p}\sqrt{\frac{\hbar Z_0 }{2\omega_0}}\left[\hat  a^{\rm R}_{\rm  in}(t)+\hat  a^{\rm R}_{\rm  out}(t)\right] + {\rm H.c.} \nonumber \,.
\end{eqnarray}
Similarly
\begin{eqnarray}\label{eq:SolutionQm1}
\hat Q_{-1}(t)&=&\sqrt{\frac{\hbar }{2\omega_{0}Z_{0}}}\left[-\hat  a^{\rm L}_{\rm  in}(t)+\hat  a^{\rm L}_{\rm  out}(t)\right] \\
&+& \frac{i}{\omega_0L_p}\sqrt{\frac{\hbar Z_0 }{2\omega_0}}\left[\hat  a^{\rm R}_{\rm  in}(t)+\hat  a^{\rm R}_{\rm  out}(t)\right]\nonumber\\
&-&\frac{i}{\omega_0L_p}\sqrt{\frac{\hbar Z_0 }{2\omega_0}}\left[\hat  a^{\rm L}_{\rm  in}(t)+\hat  a^{\rm L}_{\rm  out}(t)\right] + {\rm H.c.} \nonumber \,.
\end{eqnarray}

Our approach to find an approximative solution for this problem is the following.
We first assume that $\omega_c\equiv 1/C_{\rm s}Z_0\gg \omega_0$ and can thereby neglect the time derivatives in
Eqs.~(\ref{boundary1}) and~(\ref{boundary11}).
Within this approximation we can directly establish boundary conditions between the cavity and TL fields to be used later.
This approximation can be shown to correspond to neglecting terms $\ll \gamma$ in final equation of motion for the cavity.
This simplification leads to boundary conditions
\begin{eqnarray}
\hat Q_{-1} &=& - \hat Q \frac{ C_{\rm c}}{C+2C_{\rm c}} \label{boundary3} \\
\hat Q_{1} &=& - \hat Q \frac{ C_{\rm c}}{C+2C_{\rm c}} \label{boundary4} \,.
\end{eqnarray}
Using Eqs.~(\ref{eq:SolutionQ1}-\ref{boundary3}) and inserting $\hat Q=i\sqrt{\frac{\hbar}{2Z_{LC}}} [\hat a^\dagger(t)-\hat a(t) ]$ we get
\begin{eqnarray}
\alpha \hat a(t) &=& -\hat a_{\rm in}^{\rm L}(t)+\hat a_{\rm out}^{\rm L}(t)+\epsilon\left[ \hat a_{\rm in}^{\rm L}(t) +\hat a_{\rm out}^{\rm L}(t)\right] \nonumber \\
&+&\epsilon^*\left[ \hat a_{\rm in}^{\rm R}(t) +\hat a_{\rm out}^{\rm R}(t)\right] \\
\epsilon&=&i\frac{Z_0}{\omega_0L_{\rm b}} \\
\alpha &=& -i\frac{C_{\rm c}}{C+2C_{\rm c}}\sqrt{\frac{Z_0}{Z_{LC}}}\, \sqrt{\omega_0}\,.
\end{eqnarray}
Here we choose $Z_{LC}=\sqrt{L/(C+2C_{\rm c})}$ (and $\omega_0=1/\sqrt{L(C+2C_{\rm c})}$) since this choice
removes mixing of $\hat a$ and $\hat a^\dagger$ in the following cavity equations of motion, i.e., diagonalizes an equivalent cavity Hamiltonian.
We have also implicitly assumed that there is no mixing between annihilation and creation operators between the system and the environment, following from a rotating-wave approximation.

Analogously we establish a solution using Eq.~(\ref{boundary4}),
\begin{eqnarray}
\alpha \hat a(t) &=& -\hat a_{\rm in}^{\rm R}(t)+\hat a_{\rm out}^{\rm R}(t)+\epsilon\left[ \hat a_{\rm in}^{\rm R}(t) +\hat a_{\rm out}^{\rm R}(t)\right] \nonumber \\
&+&\epsilon^*\left[ \hat a_{\rm in}^{\rm L}(t) +\hat a_{\rm out}^{\rm L}(t)\right] \,.
\end{eqnarray}

The boundary conditions given in the main text follow a convention used in Ref.~\cite{WallsMilburn} and correspond to
redefinition of the phase of the incoming field operator as $\hat a_{\rm in}^{\rm L}\leftarrow -i \hat a_{\rm in}^{\rm L}$ and
the outgoing-field operator as $\hat a_{\rm out}^{\rm L}\leftarrow i \hat a_{\rm out}^{\rm L}$.
Similarly $\hat a_{\rm in}^{\rm R}\leftarrow -i \hat a_{\rm in}^{\rm R}$ and $\hat a_{\rm out}^{\rm R}\leftarrow i \hat a_{\rm out}^{\rm R}$.
It should be noted that in comparison to the impedance approach, we have now effectively changed the signs of out field amplitudes, changing signs of functions $s_{11}$ and $s_{12}$.
The previously-derived boundary conditions are in this notation
\begin{eqnarray}
\alpha \hat a(t) &=& \hat a_{\rm in}^{\rm L}(t)+\hat a_{\rm out}^{\rm L}(t)+i\epsilon\left[ -\hat a_{\rm in}^{\rm L}(t) +\hat a_{\rm out}^{\rm L}(t)\right]  \nonumber\\
&-&i\epsilon\left[ -\hat a_{\rm in}^{\rm R}(t) +\hat a_{\rm out}^{\rm R}(t)\right]\label{eq:bc1}  \\
\alpha \hat a(t) &=& \hat a_{\rm in}^{\rm R}(t)+\hat a_{\rm out}^{\rm R}(t)+i\epsilon\left[ -\hat a_{\rm in}^{\rm R}(t) +\hat a_{\rm out}^{\rm R}(t)\right]  \nonumber \label{eq:bc2} \\
&-&i\epsilon\left[ -\hat a_{\rm in}^{\rm L}(t) +\hat a_{\rm out}^{\rm L}(t)\right] \\
\epsilon&=&\frac{Z_0}{\omega_0L_{\rm b}} \\
\alpha &=& \frac{C_{\rm c}}{C+2C_{\rm c}}\sqrt{\frac{Z_0}{Z_{LC}}}\, \sqrt{\omega_0}=\sqrt{\gamma} \,.
\end{eqnarray}
We can also express the out-fields as a function of in-fields
\begin{align}
\hat a_{\rm out}^{\rm L}(t)&=  \alpha \hat a(t)- \frac{1}{1+2i\epsilon}\hat a_{\rm in}^{\rm L}(t)-\frac{2i\epsilon}{1+2i\epsilon}\hat a_{\rm in}^{\rm R}(t)\label{eq:CavityFanoOutaApp} \\
\hat a_{\rm out}^{\rm R}(t)&=  \alpha \hat a(t)- \frac{1}{1+2i\epsilon}\hat a_{\rm in}^{\rm R}(t)-\frac{2i\epsilon}{1+2i\epsilon}\hat a_{\rm in}^{\rm L}(t)\label{eq:CavityFanoOutbApp} \,.
\end{align}

The Heisenberg equations of motion for the cavity are (under a rotating-wave approximation and using the new notation) 
\begin{eqnarray}
\hat{\dot {a}}(t) &=& -i\omega_0 \hat a(t) +\frac{\alpha}{2}\left[ \hat a_{\rm in}^{\rm L}-\hat a_{\rm out}^{\rm L} \right]\nonumber   \\
&+& \frac{\alpha}{2}\left[ \hat a_{\rm in}^{\rm R}-\hat a_{\rm out}^{\rm R} \right] \,.
\end{eqnarray}
Inserting the solutions of Eqs.~(\ref{eq:CavityFanoOutaApp}-\ref{eq:CavityFanoOutbApp}) in the cavity equation of motion we get
\begin{align}
\hat{\dot {a}}(t) &= -i\omega_0 \hat a(t) -\alpha^2\hat a(t) +\alpha \left[ \hat a_{\rm in}^{\rm L}(t)+\hat a_{\rm in}^{\rm R}(t) \right]  \,.
\end{align}
We obtain that (in the presence of the background) the equation of motion for the cavity remains unchanged, i.e., is independent of $\epsilon$.
The interference between propagation through the cavity and parallel inductor
is described by Eqs.~(\ref{eq:CavityFanoOutaApp}-\ref{eq:CavityFanoOutbApp}).
These important results can be understood qualitatively as a consequence of that the
two transmission channels, through cavity and through the background, are in
parallel and therefore their common transmittance is the sum of the individual ones.
The only limitation of this model is therefore the validity of the single-mode treatment of the cavity (weak coupling of the cavity to the transmission lines).

\section{Fano curve tilt direction}\label{sec:TiltDirection}
An important detail of a Fano interference is the tilt direction of the Fano curve, i.e., on which side of the resonance frequency is the minimum.
The direction of the tilt is in our examples always rightwards. It is determined by two properties: (i) The change of the phase between the resonator input and output and (ii) the nature of the parallel coupling (capacitive or inductive).

In the case of a $\lambda/2$-resonator the output and input fields are related  at resonance as $b_{\rm out}(\omega_0)/a_{\rm in}(\omega_0)=(-1)^{n+1}$ for modes $n=1,2,\ldots$. The first mode $(n=1)$ then keeps the sign and the second mode $(n=2)$ inverts the sign.
The in-line resonator of Fig.~\ref{fig:setup}(b) inverts the sign at the resonance
and is then equivalent to $n=2,4,\ldots$.
In other words,  the in-line resonator and coplanar resonator are equivalent for full-wavelength modes.
However, we find that the model based on an in-line resonator can effectively describe also odd modes ($n=1,3,\ldots$), if
one makes a switch between an inductive and capacitive parallel coupling, i.e., changes the sign of $\epsilon$.
This is also the reason why our theoretical model considers explicitly a parallel inductor,
but analysis of the experiment interprets the value of $\epsilon$ to originate from a parallel capacitor.

\section{Transmission in the I-Q plane}

\begin{figure*}
\includegraphics[width=1.5\columnwidth]{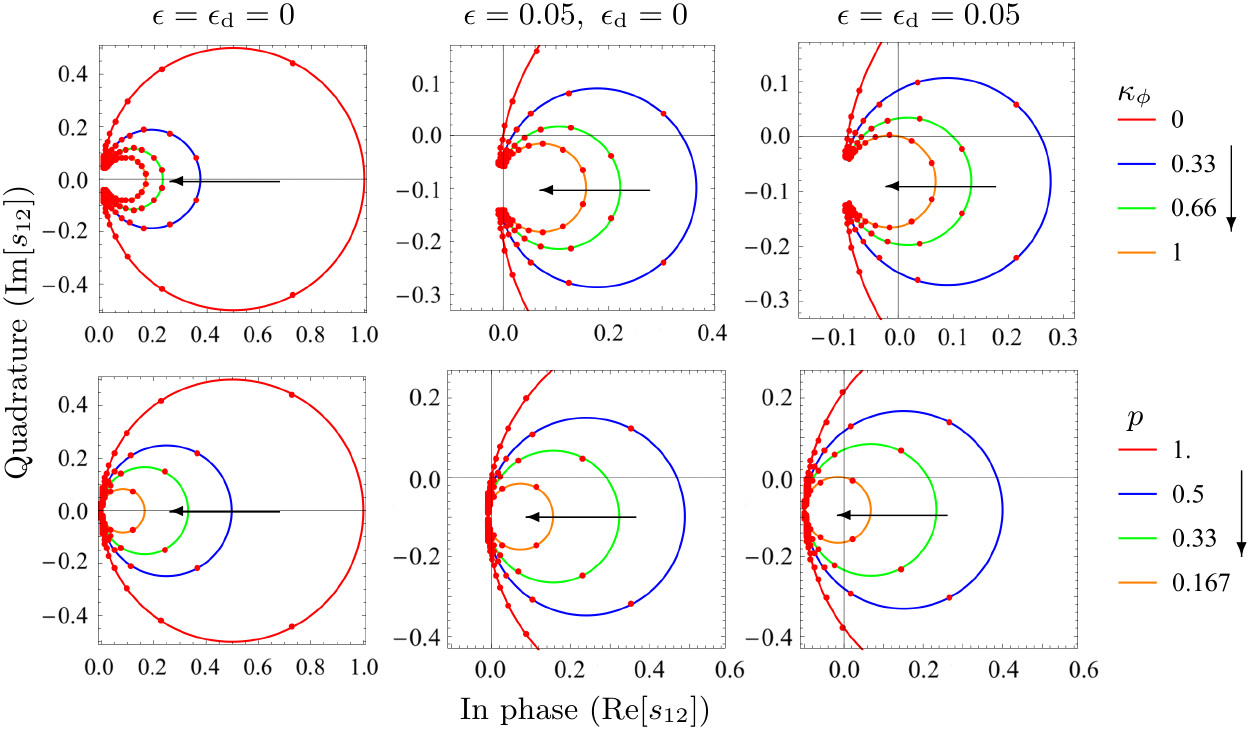}
\caption{In-phase (${\rm Re}[s_{12}]$) and quadrature (${\rm Im}[s_{12}]$) components of the transmitted field
of Figs.~\ref{fig:Theory_Heisenberg_Solution1} and~\ref{fig:ReducedPopulation}.
The frequency is sweeped from $0.95\omega_0$ to $1.05\omega_0$. Dots mark frequency differences of $1.25\gamma$.
The three top figures represent changes when increasing cavity loss rate $\kappa/2=\kappa_\phi$ (as given by the curve keys)
in three different backgrounds (titles). The three bottom figures represent changes when decreasing resonant-state population $p$ in the same backgrounds.
The radius of the transmission circles reduce when increasing $\kappa$ or decreasing $p$, as visualized by the arrows.
For $\epsilon=\epsilon_{\rm d}=0$, zero transmission [point (0,0)] appears only in the asymptotic limits
$\vert \omega-\omega_0\vert\rightarrow\infty$ or $p\rightarrow 0$. For $\epsilon=\epsilon_{\rm d}>0$
the transmission circles can overtake and touch zero transmission [point (0,0)] from below, corresponding to a dip and zero in the transmission.
}\label{fig:IQPlane}
\end{figure*}

In this paper we mainly concentrate on describing the effect of decoherence by looking at the amplitude of transmission, $\vert s_{12}\vert$.
The amplitude shows a peak (or a dip) at resonance, where the phase shifts rapidly.
In particular, at a perfect Fano dip the amplitude touches zero and the phase jumps between two values.

In this appendix, we visualize the behavior of the amplitude and the phase at the same time by representing the
previously studied line shape transformation in the I-Q plane, see Fig.~\ref{fig:IQPlane}.
Here, the I-axis corresponds to the in-phase component, i.e., to the real part of $ s_{12}$. The the Q-axis corresponds
to the quadrature component, i.e., to the imaginary part of $ s_{12}$.

For completeness, we first plot (on the left-hand side of Fig.~\ref{fig:IQPlane}) the result for the case of background-free system, $\epsilon=\epsilon_{\rm d}=0$,
where the amplitude is a Lorentzian. 
The background-free transmission forms a circle in the I-Q plane, approaching zero for $\vert \omega-\omega_0\vert\rightarrow\infty$.
For a lossy cavity, $\kappa>0$, the radius of the circle decreases below 1.
In the case of reducing probability $p$, the behavior is analogous. A difference is that
off-resonance transmission approaches faster zero for $p\rightarrow 0$.

When adding weak non-dissipative background transmission,
$0<\epsilon\ll 1$ with $\epsilon_{\rm d}=0$, the point~(0,0) (zero transmission) can be can be touched from bottom right, but only if $\kappa=0$ and $p=1$. Otherwise this point is not touched. For $\epsilon=\epsilon_{\rm d}>0$, however,
loss in the cavity transport can be balanced by a loss in the background transmission, so that
zero transmission is again reached. 
As found before, this occurs for the highest plotted $\kappa$ and lowest plotted $p$.
Here, unlike in the case of a weak non-dissipative background transmission,
the circles touch zero almost exactly from below, resulting in that the transmission dip is rather symmetric
around this point.
Similarly, in the case of strong parallel transport $\epsilon\rightarrow=\infty$, $\epsilon_{\rm d}=0$,  and $p=1$,
corresponding to the case of a conventional Fano dip as described in Sec.~\ref{sec:DissipationFreeFano} ($q=0$),
we obtain a transmission circle which has a center at (-0.5,0), touching zero exactly from below, at the resonance  $\omega=\omega_0$.




\section{Experiment details}

\begin{figure*}
\includegraphics[width=1.65\columnwidth]{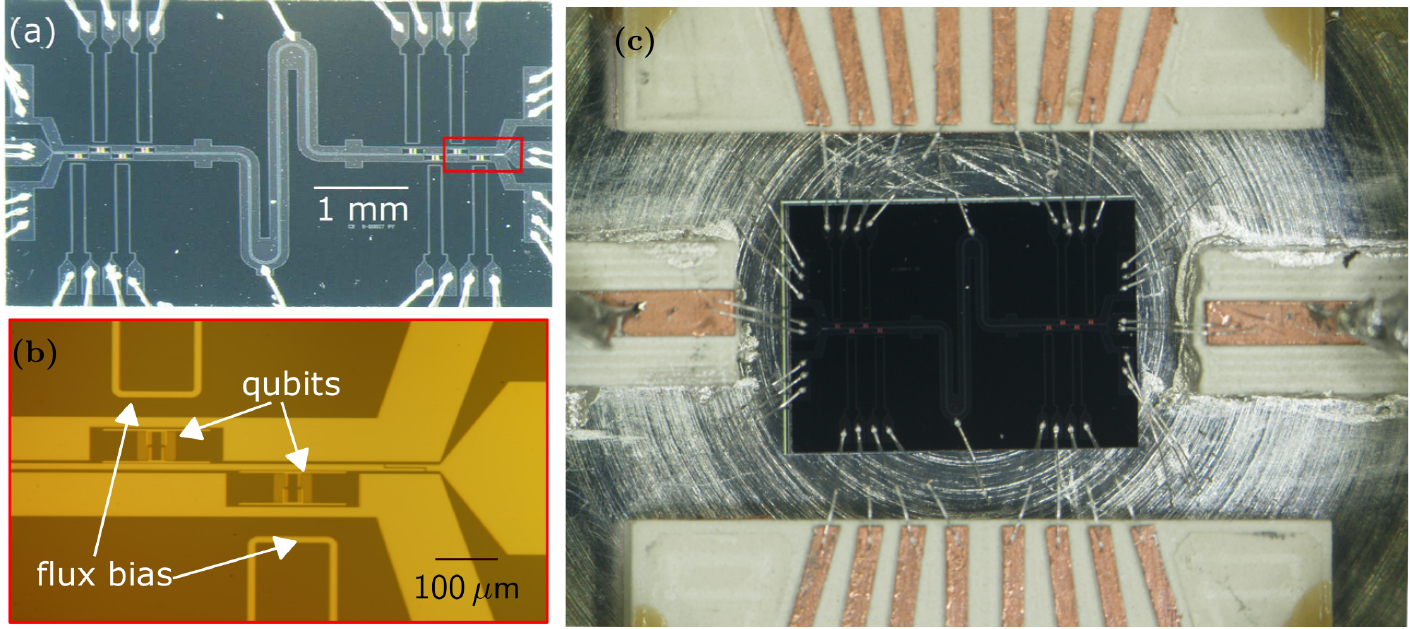}
\caption{(a) Optical micrograph of the chip
 including 8~transmon artificial atoms embedded in a driven coplanar microwave resonator.
 (b) The transmon qubits are tuned by local current lines connected to DC bias lines outside the chip.
(c) The chip inside the sample box.
}\label{fig:ExperimentGeometry}
\end{figure*}

\begin{figure*}
\includegraphics[width=1.35\columnwidth]{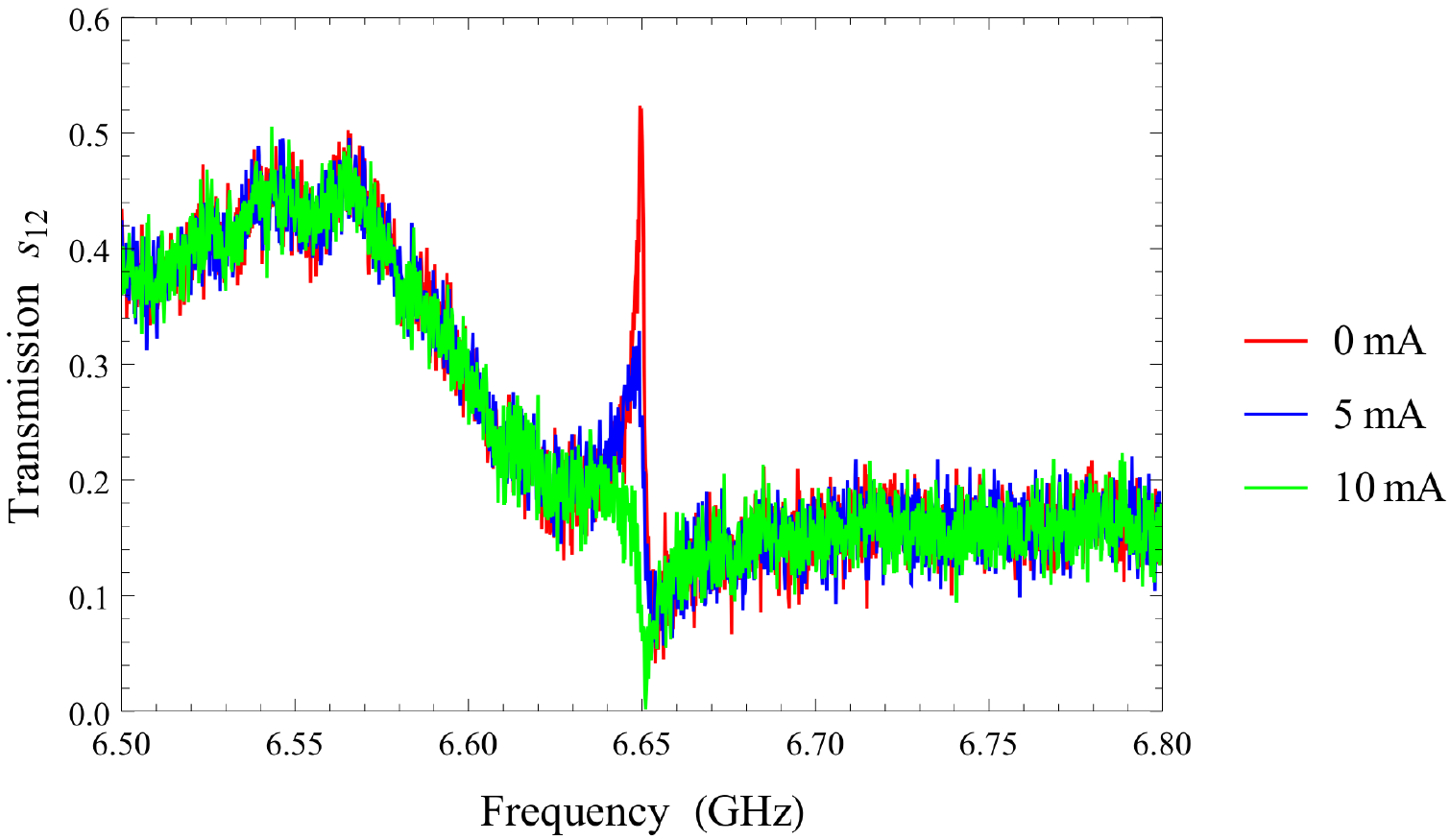}
\caption{Measured transmission $\vert s_{12} \vert$ in a broader bandwidth, drive frequency $\omega/2\pi$ varying  between 6.5~GHz and 6.8~GHz, for three values of the coil current used to tune one transmon.
The data is normalized according to high power cavity transmission, assumed to reach the theoretical maximum $\approx 0.9$
for $\epsilon-{\rm i}\epsilon_{\rm d}=0.062-0.06{\rm i}$. The maximal transmission on-resonance reduces when increasing the coil current.
}\label{fig:ExperimentBackground}
\end{figure*}

In this appendix, we give additional information to the experimental setup, the control-line layout, the background transmission,
and the used lumped-element background model.
More details of the experiment are given in Ref.~\cite{Ping2018}.

In Fig.~\ref{fig:ExperimentGeometry}(a), we show an optical micrograph of the sample.
The experimental realization includes eight transmon artificial atoms embedded in a driven coplanar microwave resonator.
The transmon artificial atoms are tuned by applying a magnetic flux across the superconducting loops of two parallel Josephson junctions, see Fig.~\ref{fig:ExperimentGeometry}(b).
The equivalent circuit model is given in Fig.~\ref{fig:setup}.
The bonding in the sample box is shown in Fig.~\ref{fig:ExperimentGeometry}(c).

The measured transmission $\vert s_{12}\vert$ is characterized by large background transmission.
In Fig.~\ref{fig:ExperimentBackground} we show wide-bandwidth transmission, between 6.5 and 6.8~GHz,
in the setup studied in Figs.~\ref{fig:Comparison1} and \ref{fig:Comparison11}.
Such background transmission appears in different cool-downs and is constant as a function of drive power.
This means that it can be modeled as a linear circuit element in transmission.
The background transmission (as well as the reduction of the cavity-transmission when increasing control current) was negligibly small in an additional experiment that included only a single transmon in the cavity.
These results imply that the Fano resonance occurs due to a crosstalk between the transmission lines and the multiple control lines.

The background transmission is not a constant as a function of frequency.
In most parts the variation is however small within the bandwidth of the resonator (within few~MHz), and can be treated as a constant in the model. The large-scale variations then correspond to different parameters in the equivalent model,
studied in Fig.~\ref{fig:Comparison1111}.
Notable is that the maximal background transmission at 6.55~GHz is of similar magnitude as maximal transmission on-resonance with the cavity. At high powers, however, the maximal (cavity-peak) transmission $\vert s_{12}\vert$ was roughly two times higher.

The background impedance can be both inductive or capacitive. In particular, it is inductive
on the left-hand side and capacitive on the right-hand side of the local maximum near 6.56~GHz (see also the discussion of the Fano tilt direction in Appendix~C).
This change reflects the appearance of a wide resonance in the background transmission around this region.
At each frequency the background transmission can however be modeled within the simple lumped-element model, having resistor $R_{\rm b}$
in series with capacitor or inductor, see Fig~\ref{fig:ExperimentBackground}. The parameters at the specifically studied point in
Fig.~\ref{fig:Comparison1111} are listed in Table~\ref{Table}.

\begin{table*}
\centering
\begin{tabular}{>{\hfil}p{65pt}<{\hfil}>{\hfil}p{65pt}<{\hfil}>{\hfil}p{65pt}<{\hfil}>{\hfil}p{45pt}<{\hfil}>{\hfil}p{45pt}<{\hfil}>{\hfil}p{45pt}<{\hfil}>{\hfil}p{45pt}<{\hfil}>{\hfil}p{45pt}<{\hfil}>{\hfil}p{45pt}<{\hfil}>{\hfil}p{45pt}<{\hfil}
}
  \hline
  \hline
$\omega/2\pi\,$(GHz) &  $R_{\rm b}\,$($\Omega$) & $1/\omega C_{\rm b}\,$($\Omega$) & $\omega L_{\rm b}\,$($\Omega$) & $\epsilon$ & $\epsilon_{\rm d} $ \\
  \hline
$6.405$ &  $0$    &  $0$     & $330$ & $-0.15 $   & $0$      \\
$6.415$ &  $10$   &  $0$     & $310$ & $-0.16 $   & $0.05$   \\
$6.46$  &  $78$   &  $0$     & $300$ & $-0.15 $   & $0.04$   \\
$6.51$  &  $160$  &  $0$     & $170$ & $-0.16 $   & $0.14$   \\
$6.56$  &  $160$  &  $0$     & $0$   & $ 0    $   & $0.31$   \\
$6.585$ &  $170$  &  $100$   & $0$   & $ 0.12 $   & $0.22$   \\
$6.66$  &  $400$  &  $420$   & $0$   & $ 0.062$   & $0.06$   \\
  \hline
  \hline
\end{tabular}
\caption{Summary of model parameters at different drive frequencies.}
\label{Table}
\end{table*}

In Fig.~\ref{fig:Comparison1111},
to remove small structure originating in the changing background transmission,
the measured data $\vert s_{12}\vert$ is modified by first removing the constant (but frequency dependent) off-resonance transmission from the on-resonance data,  and then replacing it by a background transmission
given by the parameter $\epsilon_{\rm tot}$, which was chosen to be the background transmission at the exact resonance frequency
(given in Table~\ref{Table}).
Such substitution is allowed as long as changes within the cavity bandwidth $\gamma$ stay small.
This substitution helps to better identify the transmission changes due to the cavity transmission only.


The observed capacitive coupling is supported by microwave simulations of the experimental layout,
which give a capacitive coupling $\sim 50$~fF across the multiple DC flux bias lines beside the cavity (impedance $\sim 400$~$\Omega$).
However,
the coupling from the input conductor to the bias leads does not occur through a direct coupling between the cavity and nearby conducting strips,
for which we simulate a still a relatively high value $\lesssim 3$~fF (occuring due to narrow ground conductors
in the cavity). It can however be mediated via sample box walls. In the probed frequency range,
the equivalent resistance $R_{\rm b}$ varies from zero up to the free space impedance ($377$~$\Omega$),
suggesting radiation loss to nearby conductors and/or to free space.

\end{document}